\documentclass[preprint]{aastex631}

\usepackage{amsmath}
\defcitealias{Rosborough2024ModelingNuclei}{R24}
\usepackage{multirow}
\accepted{July 15, 2025}

\begin{document}

\title{Modeling the Reverberation Response of the Broad Line Region in Active Galactic Nuclei II: Incorporating Photoionization Models}

\author[0000-0001-5055-507X]{Sara A. Rosborough}
\affiliation{George P. and Cynthia Woods Mitchell Institute for Fundamental Physics and Astronomy, Texas A\&M University, College Station TX, 77843 USA}
\affiliation{Laboratory for Multiwavelength Astrophysics, School of Physics and Astronomy, Rochester Institute of Technology,
84 Lomb Memorial Dr,
Rochester, NY 14623, USA}

\author[0000-0003-0672-2154]{Andrew Robinson}
\affiliation{Laboratory for Multiwavelength Astrophysics, School of Physics and Astronomy, Rochester Institute of Technology,
84 Lomb Memorial Dr,
Rochester, NY 14623, USA}

\author[0000-0001-7259-7043]{Triana Almeyda}
\affiliation{Department of Astronomy, University of Florida,
211 Bryant Space Science Center Stadium Road,
Gainesville, FL 32611, USA}

\author{Daniel Humphrey}
\affiliation{Department of Physics, University of Wisconsin at Madison, Madison, WI 53706, USA}
\affiliation{Department of Physics, Wheaton College, Wheaton, IL 60187, USA}

\author{Madison Noll}
\affiliation{Department of Physics, Boston University, 
590 Commonwealth Avenue, 
Boston, MA, 02215}



\begin{abstract}

The broad emission lines (BELs) emitted by Active Galactic Nuclei respond to variations in the ionizing continuum emission from the accretion disk surrounding the central supermassive black hole (SMBH). This reverberation response provides insights into the structure and dynamics of the Broad Line Region (BLR). In Rosborough et al., 2024, we introduced a new forward-modeling tool, the Broad Emission Line MApping Code (BELMAC), which simulates the velocity-resolved reverberation response of the BLR to an input light curve.  In this work, we describe a new version of \textsc{BELMAC}, which uses photoionization models to calculate the cloud luminosities for selected BELs. We investigated the reverberation responses of H$\alpha$, H$\beta$, MgII$\lambda 2800$ and CIV$\lambda 1550$ for models representing a disk-like BLR with Keplerian rotation, radiatively driven outflows, and inflows. The line responses generally provide a good indication of the respective luminosity-weighted radii. However, there are situations when the BLR exhibits a negative response to the driving continuum, causing overestimates of the luminosity-weighted radius. The virial mass derived from the models can differ dramatically from the actual SMBH mass, depending mainly on the disk inclination and velocity field. In single zone models, the BELs exhibit similar responses and profile shapes; two-zone models, such as a Keplerian disk and a biconical outflow, can reproduce observed differences between high- and low-ionization lines.  Radial flows produce asymmetric line profile shapes due to both anisotropic cloud emission and electron scattering in an inter-cloud medium. These competing attenuation effects complicate the interpretation of profile asymmetries. %248/250 words
\end{abstract}

\keywords{Reverberation mapping (2019) --- Active galactic nuclei (16) --- Supermassive black holes (1663) ---  Astronomy data modeling (1859)}


\section{Introduction} \label{sec:intro}

% BLR intro here
It is generally accepted that a supermassive black hole (SMBH) resides at the center of every large galaxy.  Active SMBHs grow by accreting matter from their surrounding disk, which releases a vast amount of electromagnetic radiation across the spectrum.  These processes power active galactic nuclei (AGN) and the accretion disk's continuum photoionizes the surrounding gas of the broad line region (BLR).  The BLR is occupied by high density ($n=10^{8}\sim10^{14} $cm$^{-3}$) gas clouds, which are located close enough to the SMBH ($<1$ pc) that the broad emission lines (BELs) they emit are Doppler broadened by $\sim2-10\times 10^3$\,km\,s$^{-1}$ \citep[e.g.,][]{Davidson1979TheObjects,Osterbrock1986Emission-LineQSOs,Netzer2008IonizedNuclei}.  Due to its proximity to the SMBH, the BLR provides a unique diagnostic of the SMBH mass ($M_\bullet$) and the physical processes that operate deep within its gravitational potential well.  In recent years, the development of infrared interferometry has made it possible to spatially resolve the BLR in some of the nearest AGN, \citep{Amorim2020The091496206,Amorim2021TheRegion,Amorim2024TheRegion}. However, this is not possible for the vast majority of AGN and alternative techniques are required to study large samples spanning a wide range of cosmological distance. One such technique is broad emission line reverberation mapping \citep[e.g.,][]{Blandford1982ReverberationQuasars,Peterson1993ReverberationNuclei,Walsh2009THECHARACTERISTICS,Rosa2015SPACETELESCOPE/i}.  

Variations in the AGN's ionizing continuum produce a corresponding response in the BELs that is modulated by light travel delays ($t^\prime$) and contains information on the physical properties of the emitting gas.  That is, the velocity-resolved response of the BELs ($L_\lambda(t,\,v^{||})$) is the convolution of the continuum light curve ($L_c(t-t^\prime)$) with a transfer function ($\Psi_\lambda(t^\prime,v^{||})$);
\begin{equation}
\label{eq:RM}
    L_\lambda(t,\,v^{||}) = \int_{-\infty}^{\infty} \Psi_\lambda(t^\prime,v^{||})L_c(t-t^\prime)\,dt^\prime
\end{equation}
where $t$ is time and $v^{||}$ is the line of sight (LOS) velocity.  The reverberation mapping technique is the time series analysis of the optical or UV light curve, used as a proxy for the driving ionizing continuum light curve, and the response of the BELs \citep{Blandford1982ReverberationQuasars,Peterson1993ReverberationNuclei}.    
% The reverberation mapping technique entails time series analysis of the response of the BELs to the driving continuum variations, with the observed optical or UV light curve serving as a proxy for the ionizing continuum. 

This technique allows the SMBH mass to be estimated for large samples of AGN.  The virial mass ($M_{\bullet,\,vir}$) can be estimated from the velocity dispersion ($\Delta v$) derived from the widths of the BEL and the size of the variable part of the BLR ($R_{BLR}$). The black hole mass ($M_\bullet$) is then, 
% size of the variable BLR

\begin{equation}\label{eq:mass}
    M_\bullet = fM_{\bullet,\,vir} = f\Delta v^2R_{BLR}/G
\end{equation}

where the ``virial scaling factor", $f$, takes into account the poorly understood structure, kinematics, and inclination of the BLR \citep[e.g.,][]{Onken2004SupermassiveNuclei}.  The size of the BLR can be obtained from reverberation mapping.  The observed continuum light curve is cross-correlated with the response (BEL) light curve, often that of H$\beta$, to obtain the response-weighted time delay, $t^\prime_{RW}$, which is assumed to provide a reasonable measure of the BLR size; $R_{BLR}\approx ct^\prime_{RW}$. BLR reverberation mapping of $\sim100$  mainly low-redshift AGN
%, a tight relationship between $t^\prime_{RW}$ and the AGN luminosity
has established a BLR radius – AGN luminosity relationship where, $t^\prime_{RW}\propto R_{BLR}\sim L_{AGN}^{0.5}$ \citep[e.g.,][]{Bentz2013THENUCLEI,Bentz2009THEAGNs}. With this relationship it is possible to determine both $R_{BLR}$ and $\Delta v$ from a single spectrum, without the need for lengthy observation campaigns, allowing the mass of SMBH to be inferred for very large samples of AGN.  To estimate the virial factor, $M_\bullet$ values determined from reverberation mapping have been compared to the empirically determined relationship between the stellar velocity dispersion of a galaxy bulge ($\sigma_\ast$) and $M_\bullet$ for nearby objects, yielding $f\approx4.5$ \citep{Woo2015THEGALAXIES,Batiste2017RecalibrationAGN}.  However, this introduces a systematic uncertainty in the derived $M_\bullet$ that is comparable to the $\sim0.4$ dex scatter in the $M_\bullet–\sigma_\ast$ relation.   

%Since the size and structure of the BLR usually cannot be found by direct imaging, we determine $f$ and $R_{BLR}$ using the reverberation mapping technique.  Based on comparisons between the $M_{\bullet}$ determined from reverberation mapping and the $M_\bullet - \sigma_\ast$ relation, $f\approx4.5$ \citep{Woo2015THEGALAXIES,Batiste2017RecalibrationAGN}.  

Although many observational and modeling studies have been conducted for the BLR, our understanding of its structure and dynamics remains incomplete.  Several reverberation mapping studies \citep[e.g., NGC 5548, NGC 3783, and NGC 3227 by][respectively]{Horne2021Space5548,Bentz2021Robotic3783,Bentz2023Velocity-resolved3227} suggest that in some AGN, the BLR has a disk-like configuration with Keplerian dynamics. But signatures of radial flows are also seen \citep[e.g., in modeling of NGC 5548 and NGC 3783 by][respectively]{Williams2020Space5548,Pancoast2018StabilityYears}, so at least one additional dynamical component appears to be required. This may be associated with a disk wind, with the relative prominence of the disk and wind components scaling with the AGN luminosity \citep{Yong2017TheModel,Elvis2017QuasarWind}. This implies that the dominant emission line producing structures in the BLR are dependent on the SMBH's accretion rate. %The existence of ``changing-look" AGN (in which the BELs disappear and/or reappear) suggests that an AGN's accretion rate, and thus the ionizing luminosity, may change dramatically over relatively short timescales ($\sim$1 year) \citep[e.g.,][]{MacLeod2019Changing-lookQuasars}.  
Furthermore, previous reverberation modeling indicates that the BLR structure and dynamics differ substantially from object to object, evolve over time in the same object \citep[e.g.,][]{Linzer2022SpectropolarimetricProducts}, and also differ for different BELs \citep[e.g.,][]{Liu2022MeasuringLines,Yu2021OzDESMonitoring,Vestergaard2006DeterminingRelationships}.  This is a concern for tracing the cosmological evolution of SMBH mass, since the vast majority of AGN spectra are obtained in the optical band (e.g., from the Sloan Digital Sky Survey (SDSS) \citet{Shen2011A7} and the Swift BAT AGN Spectroscopic Survey (BASS) catalog \citet{Koss2022BASS.Data}).  Therefore, careful modeling of reverberation mapping data for multiple BELs is necessary for tight constraints on $R_{BLR}$, $f$, and $M_\bullet$ \citep[e.g.,][]{Assef2011BLACKLINES,McLure2004TheMasses}. 

The comparatively few reverberation mapping studies that include high and low ionization broad lines have reported different profile shapes in the same AGN \citep[e.g., NGC 5548][]{Kriss2019SpaceSpectrum,Pei2017Space5548}, as have more numerous single-epoch studies \citep{Richards2011UNIFICATIONEMISSION,Espey1989H-alphaQuasars}.  Differences in profile shapes have also been observed within AGN datasets \citep[e.g.,][]{Oh2022BASS.Demographics, Bentz2010THELINES}.  High-ionization lines (HILs), such as HeII, NV, OVI, Ly$\alpha$, and CIV, are broader than low-ionization lines (LILs), tend to be blue-asymmetric, and blueshifted with respect to LILs. The HILs also tend to have shorter reverberation mapping time delays.  Therefore, it is thought that HILs originate from outflowing gas close to the central engine \citep{Richards2011UNIFICATIONEMISSION,Gaskell1982AMotions}.  In contrast, LILs, such as HeI, MgII, H$\alpha$, and H$\beta$ have more symmetric profiles with shorter time delays in the wings rather than the core of the profile, suggesting a disk structure with virial gas motion \citep[e.g.,][]{Yong2017TheModel}.  Biconical winds have also been invoked to explain why some AGN exhibit broad absorption lines (BALs) or narrow absorption lines (NALs), while most have neither, possibly as a consequence of viewing angle \citep{Baskin2018DustNuclei,Gallagher2015InvestigatingQuasars,Elvis2000AQuasars}.  For the reasons outlined above, it is essential to establish precise inter-calibrations between the various emission lines for accurate SMBH mass determinations.

In \citet{Rosborough2024ModelingNuclei}, henceforth \citetalias{Rosborough2024ModelingNuclei}, we introduced a BLR reverberation mapping code, \textsc{Belmac}.  The models presented in \citetalias{Rosborough2024ModelingNuclei} used hydrogen recombination theory to approximate the line emission. This served as a foundation for exploring the dependence of the line response functions and line profiles on the geometry and velocity field parameters of the BLR.  However, it is necessary to use detailed photoionization models to more accurately simulate the responses of the BELs. 
Several authors have used large grids of photoionization models with BLR models involving either discrete cloud distributions or hydrodynamic flows to investigate the structure and dynamics of the BLR \citep[e.g.,][]{Korista2000Locally5548, Korista2004WhatNuclei, Matthews2020StratifiedProperties}, or the reverberation response of various BELs \citep[e.g.,][]{Goad1993ResponseNuclei,OBrien1994ResponseEmission,OBrien1995TheNuclei,Kaspi2000REVERBERATIONNUCLEI, Horne2004ObservationalMapping,Goad2012TheTorus,Mangham2019DoRegion}.
%\textbf{Simulations of multiple BELs have been performed to investigate the structure and dynamics of the BLR using photoionization codes by \citep[e.g.,][]{Korista2000Locally5548, Korista2004WhatNuclei, Matthews2020StratifiedProperties}, and for reverberation mapping by \citep[e.g.,][]{Goad1993ResponseNuclei,OBrien1994ResponseEmission,OBrien1995TheNuclei,Kaspi2000REVERBERATIONNUCLEI, Horne2004ObservationalMapping,Goad2012TheTorus,Mangham2019DoRegion}.} 
In the spirit of this previous work, we here present a new version of \textsc{Belmac}, which utilizes grids of photoionization models to calculate the cloud emission of various selected lines.  We conduct a parameter exploration of velocity-resolved responses to a square-wave continuum pulse for H$\alpha$, H$\beta$, CIV$\lambda\lambda$ 1550, 1548\,\AA, and MgII$\lambda\lambda$ 2795, 2803\,\AA.  We consider 3 configurations for the BLR cloud ensemble: a rotating disk, a disk with a radial flow, and a biconical radial flow.

In Section~\ref{sec:meth} we provide an overview of \textsc{Belmac} which includes a new bicone geometry configuration, and describe the photoionization model grids.
We present response functions and line profiles for three BLR configurations in Section~\ref{sec:results}.  Our results are discussed in Section~\ref{sec:dis}, including an example of a 2-zone, disk-wind model.  Lastly, we summarize our results in Section~\ref{sec:concl}.   

\section{Overview of BELMAC} \label{sec:meth}

In \textsc{Belmac}, the BLR is represented as a 3D ensemble of discrete clouds that are randomly distributed within a defined structure. The geometry setup is similar to that used in the torus reverberation mapping code \textsc{Tormac} \citep{Almeyda2017ModelingIllumination}, but here the clouds are dust-free gas clouds. These are photoionized by the ionizing continuum emission from the accretion disk, which is assumed to be a point source from the perspective of the BLR. The clouds are distributed within 3 basic configurations: a sphere, a flared disk, or a bicone, which can be combined (e.g., a disk with a biconical wind). A velocity field is also specified, either Keplerian rotation, radial flow, random turbulence, or any combination thereof. Each cloud is assigned a velocity vector based on the velocity field from which the LOS velocity is calculated. The user provides the AGN's bolometric luminosity, $L_{AGN}$, and the ionizing photon luminosity, $Q_H$, calculated from the spectral energy distribution (SED). A ``driving" input light curve must also be supplied to represent the temporal variability of the continuum luminosity. This may be an observed optical or UV light curve serving as a proxy for the accretion disk's ionizing continuum light curve.

%An input light curve must also be provided, which represents the variability of the  AGN's ionizing luminosity.}

%In \textsc{Belmac}, the BLR is represented as a 3D ensemble of discrete clouds that are randomly distributed within a defined structure. The geometry setup follows a similar approach to that used in \textsc{Tormac} developed by \citet{Almeyda2017ModelingIllumination}, but the clouds within the ensemble are dust-free gas clouds. A velocity field is also specified, and each cloud is assigned a velocity vector.  The cloud positions are randomly generated, and the LOS velocity of each cloud is calculated based on the type of velocity field; Keplerian, radial flow, or turbulent.  The BLR clouds can be arranged in 3 basic configurations; a sphere, flared disk, or bicone, which can be combined (e.g., a disk with a biconical wind).  The gas is photoionized by the ionizing continuum emission from the accretion disk, which is assumed to be a point source from the perspective of the BLR.  The user provides the AGN's bolometric luminosity, $L_{AGN}$, and the spectral energy distribution (SED). \textcolor{purple}{An input light curve must also be provided, which represents the variability of the  AGN's ionizing luminosity.}%\textcolor{teal}{Additionally, the user provides the ``driving" UV/optical continuum light curve as a proxy for the  accretion disk's ionizing continuum light curve.}  
Following the initial cloud ensemble setup, the luminosities of the selected lines are computed as a function of LOS velocity and time by integrating over the cloud ensemble at each observer time step, taking into account light-travel delays.  Table~\ref{tab:para} summarizes the parameters of \textsc{Belmac} with their respective descriptions and symbols, as well as the values of the AGN properties and the parameters used for the BLR models presented in this paper.

\begin{table}
\centering
\caption{BLR geometry and property parameters used to define a model and ionizing field properties calculated from the SED used in this paper.  The bold text indicates free parameters. }
\begin{tabular}{l|c|l}
\label{tab:para}
    Parameter Description & Symbol & Model Values\\
    \hline\hline
    Bolometric luminosity  & $L_{AGN}$ & $10^{45}$\,[erg\,s$^{-1}$] \\
    Total ionizing flux & $Q_H$ & $1.27\times10^{55}$\,[photons\,s$^{-1}$]\\
    Ionizing luminosity & $L_{ion}$ & $5.72\times10^{44}$\,[erg\,s$^{-1}$]\\
    Mean ionizing photon energy & $h\nu_{ion}$ & $28.06$\,[eV\,photon$^{-1}$]\\
    Outer radius & $R_d$ & $1.23\times10^{18}$\,[cm]\\
    \textbf{SMBH mass} & $M_\bullet$ & $10^8$\,[M$_\sun$]\\
    \textbf{Scaled radial size} & $Y_{BLR}$ & 20 \\
    \textbf{Angular half-width} & $\sigma$ & $20^\circ$ \& $10^\circ$\\
    \textbf{Half-opening angle} & $\mathcal{O}$ & $80^\circ$ \& $60^\circ$\\
    \textbf{Gas density in a cloud at $R_d$} & $n(R_d)$ & $10^{9.5}$\,[cm$^{-3}$]\\
    \textbf{Cloud density power-law index} & $p$ & 0 \& 2\\
    \textbf{Inclination to observer} & $i$ & $0^\circ,~30^\circ,$ \& 60$^\circ$\\
    \textbf{Zonal covering fraction} &  $C_{f,zone}$ & 0.3 \\
    \textbf{Softening parameter} & $s_p$ & 1\\
    \textbf{Initial velocity} & $v_o^2$ & $v_{Kep}$ \& $\sqrt{2} v_{Kep}$\\
    \textbf{Gas density power-law index}  & $s$ & 0 \& $-2$\\
    \textbf{Total number of clouds}\footnote{Only impacts statistical noise of the model.} & $N_{tot}$ & $5\times10^5$\\
    Emission line fraction & $elf$ & $0-1$ \\
    \hline
    Gas density range \footnote{Listed ranges are the largest of any model. Ranges vary with $s$ and $p$ parameters.} & $n(R_{in})-n(R_d)$ & $1.3\times10^{12}-10^{9.5}\,$cm$^{-3}$\\
    Column density range & $N_c(R_{in})-N_c(R_d)$ & $3.8\times10^{24}-10^{22}\,$cm$^{-2}$\\
    Ionization parameter range & $U(R_{in})-U(R_d)$ & $3.7-10^{-2}$ \\
\end{tabular}
\end{table}

%The BEL MApping Code (\textsc{Belmac}) is a reverberation mapping modeling tool that computes the light curve and time-series spectral response of the BLR emission to variations in the AGN continuum for a 3D ensemble of discrete clouds, that are randomly distributed within an enclosed structure \citepalias{Rosborough2024ModelingNuclei}.  The code builds on the TOrus Reverberation MApping Code (TORMAC), by \citet{Almeyda2017ModelingIllumination}, which simulates the infrared reverberation response from the circumnuclear dusty torus just beyond the BLR. 

\subsection{General Summary of BELMAC} \label{sec:meth:pre}

The original version of \textsc{Belmac} is described in detail in \citetalias{Rosborough2024ModelingNuclei}; here we provide a brief summary of its main features. The clouds are distributed within an overall geometry that is modeled as a flared disk. The disk is characterized by the half-angular width $\sigma$, and the inclination of its axis to the observer's LOS, which ranges from $i=0\degr$ (face-on) to $i=90\degr$ (edge-on). Setting $\sigma = 90\degr$ produces a spherical cloud distribution. 

The outer radius of the BLR is set by the dust sublimation radius $R_d$,

\begin{equation} \label{eq:Rd}
    R_d = 0.4\left(\frac{L_{AGN}}{10^{45}\mathrm{erg\,s}^{-1}}\right)^{1/2}\left(\frac{T_{sub}}{1500\,\mathrm{K}}\right)^{2.6}\mathrm{\,pc}
\end{equation}

where $T_{sub}$ is the dust sublimation temperature \citep{Nenkova2008AGNMEDIA}.  The inner radius, $R_{in}$, is controlled by the scaled size $Y_{BLR} = R_d/R_{in}$.  The radial cloud distribution is a power law, $N(r)dr\propto (r/R_d)^p$dr, where $p$ is a free parameter \citep[e.g.,][]{Rees1989SmallNuclei}.  Since the ensemble is a representative sub-sample of $\sim 5\times10^7$ clouds \citep[e.g.,][]{Arav1998Are4151,Dietrich1999Structure273,Muller2020RadiationDisks}, the user may specify the number of clouds in the ensemble to alter the trade-off between statistical noise and run-time, without affecting any BLR properties.  Within the defined geometric structure, the clouds are randomly assigned positions in spherical coordinates $(r,~\theta,~\phi)$, where $r$ is the radial distance, $\theta$ is the polar angle, and $\phi$ is the azimuthal angle.  The disk can either have ``sharp'' edges where $\cos\theta$ is uniformly distributed between $90\degr\pm\sigma$ or ``fuzzy" edges where $\theta$ is drawn from a Gaussian distribution in the complementary angle, ($90\degr-\theta$).  Following \citet{Rees1989SmallNuclei,Netzer2008IonizedNuclei} the clouds have a gas density, $n(r)$, which varies radially as a power-law with index $s$, $n(r) \propto r^s$.  The clouds are assumed to be spherical, and the cloud mass is constant. Therefore, the size of each cloud is $R_{cl}(r)=R_{cl}(R_d)(r/R_d)^{-s/3}$, where $R_{cl}(R_d)\propto \sqrt{C_f}$, and $C_f$ is the integrated covering fraction. $C_f$, $n(R_d)$, and the index $s$ are free parameters.  For our chosen parameter values, the ranges of $U,~N_c$, and $n$ are listed in Table~\ref{tab:para}. These ranges are similar to typical BLR values inferred from observations and photoionization modeling \citep[e.g.,][]{Davidson1979TheObjects,Osterbrock1986Emission-LineQSOs,Netzer2013TheNuclei}.  

%the averages, weighted by the radial cloud distribution, are $\bar{n}=1.3\times10^{11}\,$cm$^{-3}$, $\bar{N_c}=5.4\times10^{23}\,$cm$^{-2}$, and $\bar{U}=0.006$ including all model combinations of $s$ and $p$.

%\textbf{For our chosen parameters of the cloud properties listed in Table~\ref{tab:para}, $10^{9.5}\lesssim \,n(r)\,\lesssim10^{12}\,$cm$^{-3}$ and $10^{22}\lesssim \,N_c(r)\,\lesssim10^{24}\,$cm$^{-2}$. Given $Q_H$ from the SED, the ionization parameter is constant at $9.2\times10^{-3}$ when $s=-2$.  When $s=0$, $U(r)$ increases to $3.6$ at the inner radius, which is higher than the values of the typical BLR ionization parameter, $\sim0.1-0.01$ \citep[e.g.,][]{}.  However, for the majority of the model BLR $0\lesssim\,U(r)\,\lesssim10^{-2}$.}            

There are 3 types of velocity fields incorporated in \textsc{Belmac}; radial flows, Keplerian orbits, and random motion to represent macroturbulence.  These velocity fields can be used either individually or in combination. %For instance, \textsc{Belmac} has the capability to simulate scenarios such as a rotating disk with a radial component, an outflow with turbulence, or a combination of rotation with radial flows and turbulence. In \citetalias{Rosborough2024ModelingNuclei}, we focused on BLR models that separately explored radial outflows and Keplerian motion.

The radial acceleration, $a_{rad}$, is assumed to be determined by the gravity and radiation pressure of SMBH.  The equation of motion for a cloud is,  

 \begin{equation} \label{eq:eom}
     a_{rad} = \frac{GM_\bullet}{r^2} \left(\Gamma F_M -1 \right)
 \end{equation}
 
where $\Gamma=L_{AGN}/L_{Edd}$, is the Eddington ratio, and $F_M$ is the force multiplier.  The force multiplier is given by $F_M=\alpha_f/\sigma_TN_c$, where $\sigma_T$ is the Thomson cross-section, $N_c(r)$ is the cloud's column density, and $\alpha_f$ the fraction of $L_{AGN}$ that is absorbed by the cloud. The latter quantity is also dependent on $N_c(r)$ as well as the ionization parameter, $U(r,t)$ \citep{Marconi2008TheGalaxies,Netzer2008IonizedNuclei,Netzer2010THENUCLEI}. When $\Gamma F_M>1$ the cloud is accelerated outward and when $\Gamma F_M<1$, it is accelerated inward.  As the BLR velocity dispersion $\Delta v << c$, the clouds are assumed to be static over timescales of interest and therefore \textsc{Belmac} does not calculate the cloud's trajectories, however, it does calculate the percentage of clouds accelerating and decelerating, according to the sign of Equation~\ref{eq:eom}.       

For the Keplerian velocity field circular orbits are assumed $v_{rot}\propto (r/R_d)^{-1/2}$, and the cloud orbits are randomly inclined within the range $90\degr-\sigma \leq |\theta| \leq 90\degr+\sigma$.  The directions of the turbulent velocity components for each cloud are randomly drawn from a Gaussian distribution and the magnitude scales as $r^{-1/2}$ \citep[e.g.,][]{Perez1992TheFunctions}.  Figure 5 of \citetalias{Rosborough2024ModelingNuclei} shows examples of the $v(r)$ profiles for Keplerian and outflow velocity fields using parameters similar to those used in this paper.  For a single BLR geometry and velocity field, there are 12 free parameters in total, which are listed in Table~\ref{tab:para} with the values for the BLR models considered here.       

In \citetalias{Rosborough2024ModelingNuclei}, the cloud luminosity was calculated using hydrogen recombination theory.  In this case, cloud luminosity scales linearly with the ionization parameter, $L_{cl}\propto U(r,t^\prime)$, until the cloud becomes fully ionized, when $L_{cl}=\mathrm{constant}$. The ionization parameter is given by $U(r,t^\prime)= Q_H(t^\prime)/4\pi r^2cn(r) \propto Q_H(t^\prime)( r/R_d)^{-s-2}$, where the time dependence is given by the input light curve and the initial total ionizing photon luminosity, $Q_H(t^\prime=0)$, is determined from the AGN's SED. 
%In order to determine cloud luminosity with time, $U(r,t^\prime)\propto Q_H(t^\prime)(r/R_d)^{-s-2}$ and $U(R_d,0)=Q_H(0)/4\pi R_d^2cn(R_d)$.  

At the high gas densities characteristic of the BLR, the clouds are expected to become optically thick in certain lines, including H$\alpha$ and H$\beta$. They will therefore emit anisotropically in those lines, as photons escape more readily from the face of the cloud that is illuminated by the central continuum \citep[e.g.,][and references therein]{Ferland1992AnisotropicNuclei,OBrien1994ResponseEmission, Gondhalekar1996TheStudies}. This effect, anisotropic cloud emission (ACE), is approximated using the emission line fraction ($elf$), which is the fraction of the illuminated cloud surface the observer sees, $elf=1/\pi\arccos(\sin\theta\cos\phi)$.  When an observer views the non-illuminated face of a cloud, $elf=0$, and when they view the fully illuminated face, $elf=1$. %This feature may be turned off to simulate isotropic cloud emission (ICE).  
The effects of isotropic and anisotropic cloud emission were presented in \citetalias{Rosborough2024ModelingNuclei}.  ACE effects on the velocity-resolved response have also been previously discussed by \citet{OBrien1994ResponseEmission}.  A visualization of ACE modeled by $elf$ is shown in Figure~\ref{fig:geo}. 

% accretion disk illumination (ADI) 
The BLR itself could be anisotropically illuminated because of limb darkening of the accretion disk.  In this case, the incident ionizing flux varies with the polar angle $F_{AGN}\propto 1/3\cos\theta(1+2\cos\theta)$ \citep{Almeyda2017DustyAGN,Netzer1987QuasarRegion}, and therefore so do $R_d$ and $U(r,t^\prime)$.  The degree of anisotropic illumination is specified by the softening parameter, $s_p$, where $s_p<1$ for anisotropic and $s_p=1$ for isotropic BLR illumination.  The $L_{AGN}$ in Equation~\ref{eq:Rd} is replaced with    

\begin{equation}
\label{eq:adi}
    L(\theta)=[s_p+1/3(1-s_p)\cos\theta(1+2\cos\theta)]L_{AGN}
\end{equation}

and the outer radius of the BLR, now $R_d(\theta)$, has a figure-8 shape that is similar to the BLR picture shown in Figure 13 of \citet{Baskin2018DustNuclei}.  

\subsection{New Capabilities in BELMAC} \label{sec:meth:new}

\subsubsection{Bicone Geometry} \label{sec:meth:geo}

\begin{figure}
    \centering
    \includegraphics[width=0.7\columnwidth]{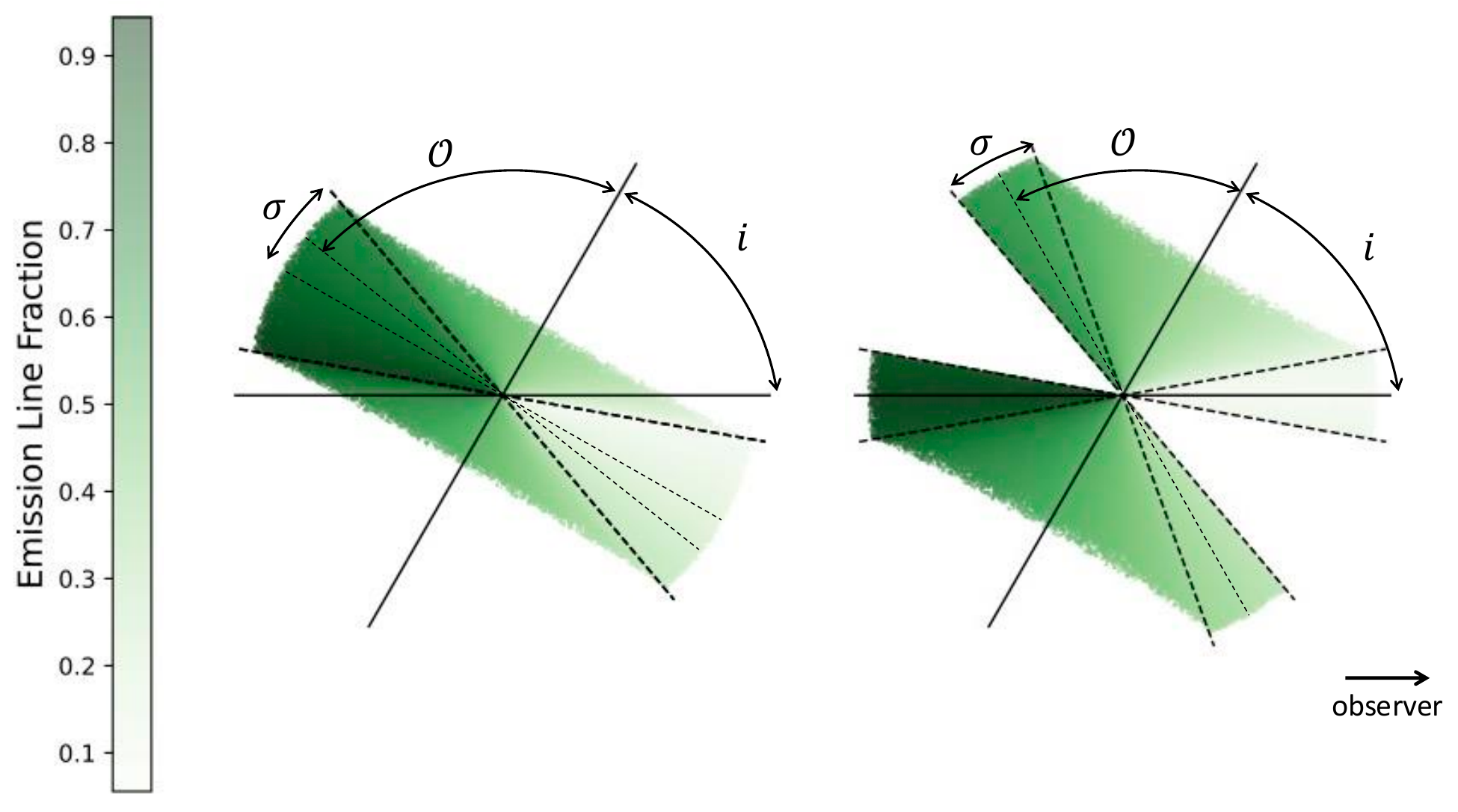}
    \caption{Cross-sections of example 3D BLR cloud ensemble geometries, a disk (left) and a bicone (right).  The shading indicates the fraction of the illuminated face of the cloud that is viewed by the observer; darker shades indicating a larger fraction.  The inclination ($i$), cone half-opening angle ($\mathcal{O}$), and angular width ($\sigma$) are labeled. The horizontal solid line is the observer's LOS, who is to the right.  The solid line inclined by $i$ is the BLR's polar axis.  \emph{Left:} The disk configuration is defined by setting $\mathcal{O}=90^\circ-\sigma/2$, where $\sigma$ is the half-angular width of the disk. \emph{Right:} A hollow biconial BLR with $\mathcal{O}=i$, such that the observer's LOS bisects the ``walls'' of each cone.  All geometry configurations are manipulations of a bicone, which allows for several possible BLR structures.
    }
    \label{fig:geo}
\end{figure}
 
We have implemented a more flexible geometry set-up, where the basic configuration is a hollow bicone that can be manipulated to create other structures, such as a sphere or flared disk.  Different geometry configurations can be combined to create more complex BLR structures, such as a disk with a bicone.  We will refer to an individual structural component of the BLR as a `zone'.  The shape of a zone is defined by two parameters: angular width $\sigma$, and half-opening angle $\mathcal{O}$. Depicted in Figure~\ref{fig:geo} are cross-sections of an inclined disk and hollow bicone. Referring to the Figure, $\sigma$ is the angle between the inner and outer surfaces of the cone and $\mathcal{O}$ is the angle between the polar axis and the bisector of $\sigma$. The disk and sphere geometries are special cases of the basic bicone. To create a disk, $\mathcal{O} = 90^\circ-\sigma/2$, and to create a sphere, $\sigma = 90^\circ$, $\mathcal{O} = 45^\circ$. Therefore, in the disk configuration, $\sigma$ is the half-angular width of the disk. The inclination, $i$, is the angle between the polar axis (i.e., the symmetry axis) and the observer's LOS. Therefore, when $i=0^o$, the system is viewed along the polar axis, and for $i=90^o$ it is viewed perpendicular to this axis (i.e., side-on).  %For the bicone in Figure~\ref{fig:geo}, the half-opening angle is equal to the inclination, such that the observer's LOS bisects the interstitial sectors between the ``walls" of near- and far-side cones. 

The bicone and disk geometries may be combined. In this case, they are treated as separate zones with their own geometry, cloud properties, and velocity fields. Thus, each zone is described by its own values of $Y_{BLR}$, $\sigma$, $\mathcal{O}$, $s$, $p$, $n(R_d)$ and covering fraction ($C_{f,\,zone}$).  We assume that the total covering fraction of the BLR, $C_f$, is the sum of the disk and bicone's respective covering fractions.  Since each zone may also have a different $Y_{BLR}$, but the same $R_{in}$, one zone may extend beyond $R_d$.  For a 2-zone BLR model of a disk and bicone with different velocity fields, there are at least 18 free parameters (see Table~\ref{tab:para}).  

%the total number of clouds is doubled, as, the disk and bicone will each have the number of clouds specified.  Doubling the number of clouds does not affect the integrated covering fraction nor the sizes of the clouds because our ensemble is a representative sub-sample of a BLR with $10^7$ clouds.  Therefore, for an ensemble of 50,000 clouds, each cloud represents 200 clouds.    

\subsubsection{Cloud Emission and Photoionization Model Grids} \label{sec:meth:pims}

The models computed with the simple reprocessing approximation described in \citetalias{Rosborough2024ModelingNuclei} were sufficient to explore the dependence of the transfer functions and line profiles on the parameters governing the cloud distribution, geometry, and velocity field.  However, while this may be a reasonable first approximation for H$\alpha$, and other Balmer lines, line formation processes in the high density environment of the BLR are more complex than hydrogen recombination theory. Furthermore, this approach cannot accurately model the behavior of collisionally excited lines of metal ions such as CIV$\lambda\,1548,1550$ or MgII$\lambda\,2800$, for which the ionization structure and thermal balance of the clouds, as well as radiative transfer, are important.

Therefore, we have implemented the capability to use the output of photoionization models to calculate emission line fluxes. \textsc{Belmac} reads multi-dimensional grids of pre-computed  line fluxes, covering wide ranges in $U$, $n$, and $N_c$, for a selected set of emission lines.  The photoionization model grids used for the \textsc{Belmac} models presented in this paper were created with \textsc{Cloudy v23} \citep{Gunasekera2023TheCloudy}, however, any photoionization modeling code can be used. 
   
In order to create the \textsc{Cloudy} photoionization model grids, we use plane-parallel geometry and assume the standard set of Solar abundances \citep{AllendePrieto2002ARatio,AllendePrieto2001TheSun,Holweger2001PhotosphericImplications,Grevesse1998StandardComposition}. We use \textsc{Cloudy}'s ``intensity case'', that is, the energy flux incident on the cloud's illuminated surface is specified by the ionization parameter ($U$).  The grid covers a wide range in $U$ that encompasses empirical estimates for the BLR and allows a wide range in distance from the ionizing source ($r$), $10^{-5} \leq U \leq 10$ with a step size of 0.25 dex.  Similarly, column density and gas density have broad ranges, $10^{19} \leq N_c \leq 10^{25}$\,cm$^{-2}$ and $10^{7.5} \leq n \leq 10^{15}$\,cm$^{-3}$, with 1.0 dex and 0.5 dex intervals, respectively.

It is also necessary to specify an SED for the incident continuum.  Ideally, to use \textsc{Belmac} for modeling AGN reverberation mapping data, a unique grid computed for the AGN's SED would be used.  In practice, we have created 3 sets of photoionization grids using 3 SEDs constructed by \citet{Jin2012ASpectra}, that represent AGN with $\Gamma=0.10,~1,$ and 10.  The 3 SEDs are included with \textsc{Cloudy v23}.  For the BLR models presented in Section~\ref{sec:results}, we adopt \citet{Jin2012ASpectra}'s SED for $\Gamma\approx0.1$, $L_{AGN}=10^{45}$\,erg\,s$^{-1}$, and $M_\bullet=10^8$\,M$_\odot$ (see Table~\ref{tab:para} for other SED-related quantities).  The computed photoionization model grids include the following BELs: Ly$\alpha\lambda$1215, NV$\lambda\lambda$1238,1242, SiIV$\lambda\lambda$1394,1403, CIV$\lambda\lambda$1550,1548, HeII$\lambda$1640, CIII]$\lambda\lambda$1906,1908, MgII$\lambda\lambda$2795,2802, HeII$\lambda$4685, H$\beta\lambda$4861, HeI$\lambda$5875, H$\alpha\lambda$6562.  For the purposes of this paper, we will focus on the responses of H$\alpha$ and H$\beta$, which are the BELs most commonly used for reverberation mapping, and the high and low ionization lines, CIV and MgII, respectively. For the CIV and MgII doublets, we use the sum of the line fluxes. 

\textsc{Cloudy} calculates both the total flux $F_{\lambda,\,tot}$ emitted per unit area by the cloud and the flux that escapes the illuminated face of the cloud, the inward flux, $F_{\lambda,\,in}$.  Even if clouds are not completely ionized, some emission can escape the non-illuminated side -- the outward flux, $F_{\lambda,\,tot} - F_{\lambda,\,in}$. To find the flux that is directed towards the observer, the inward and outward fluxes are weighted by \emph{elf}, which describes a cloud's illuminated fraction (Section~\ref{sec:meth:pre}),

\begin{equation}\label{eq:flux}
    F_{\lambda} = (elf)F_{\lambda,\,in} + (1-elf)(F_{\lambda,\,tot} - F_{\lambda,\,in}).
\end{equation}

\begin{figure}[h!]
    \centering
    \includegraphics[width=0.8\textwidth]{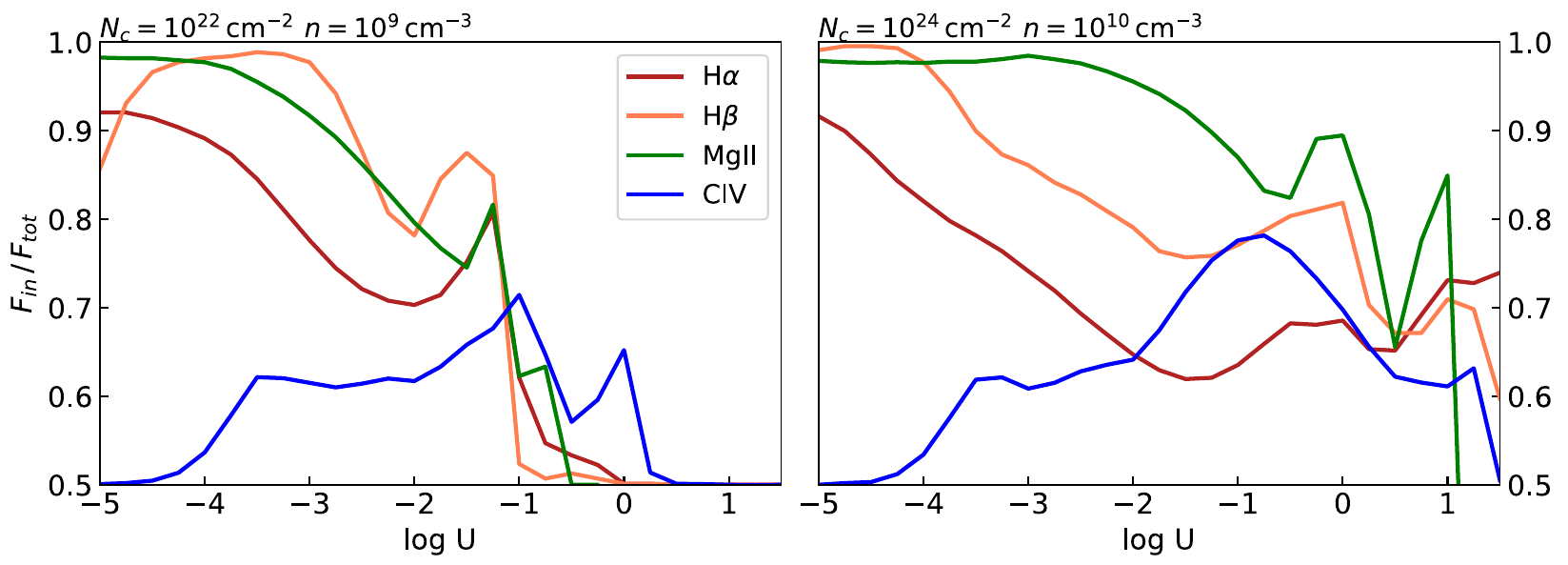}
    \caption{The variation of the inward to total flux ratio with $U$ (0.25 dex intervals) for clouds with $n=10^{9}$\,cm$^{-3}$, $N_c=10^{22}$\,cm$^{-2}$ (left) and $n=10^{10}$\,cm$^{-3}$, $N_c=10^{24}$\,cm$^{-2}$ (right).  The emission line fluxes of CIV (blue), MgII (green), H$\beta$ (orange), and H$\alpha$ (red) were calculated with \textsc{Cloudy}.  In the left panel, MgII has $F_{tot} = 0$ at $\log\,U> -0.5$ (see Figure~\ref{fig:FvU}), while the other BELs become optically thin at high $U$ values.  A cloud is considered optically thin in an emission line when $F_{\lambda,\,in}/F_{\lambda,\,tot}=0.5$.}
    \label{fig:FRvU}
\end{figure}

The fraction of flux that is directed inwards and outwards depends on the BEL, $U$, $n$, and $N_c$. Figure~\ref{fig:FRvU} shows the variation of $F_{\lambda,\,in}/F_{\lambda,\,tot}$ with $U$ for each BEL and two example combinations of the gas density and the column density; $n=10^{9}$\,cm$^{-3}$, $N_c=10^{22}$\,cm$^{-2}$ and $n=10^{10}$\,cm$^{-3}$, $N_c=10^{24}$\,cm$^{-2}$.  For each cloud, given its $U$, $n$, and $N_c$ values, \textsc{Belmac} linearly interpolates the photoionization model grids to determine the corresponding values of $F_{\lambda,\,tot}$ and $F_{\lambda,\,in}$ and calculates the emission line fluxes for the desired lines according to Equation~\ref{eq:flux}. 

\begin{figure}[h!]
    \centering
    \includegraphics[width=0.8\textwidth]{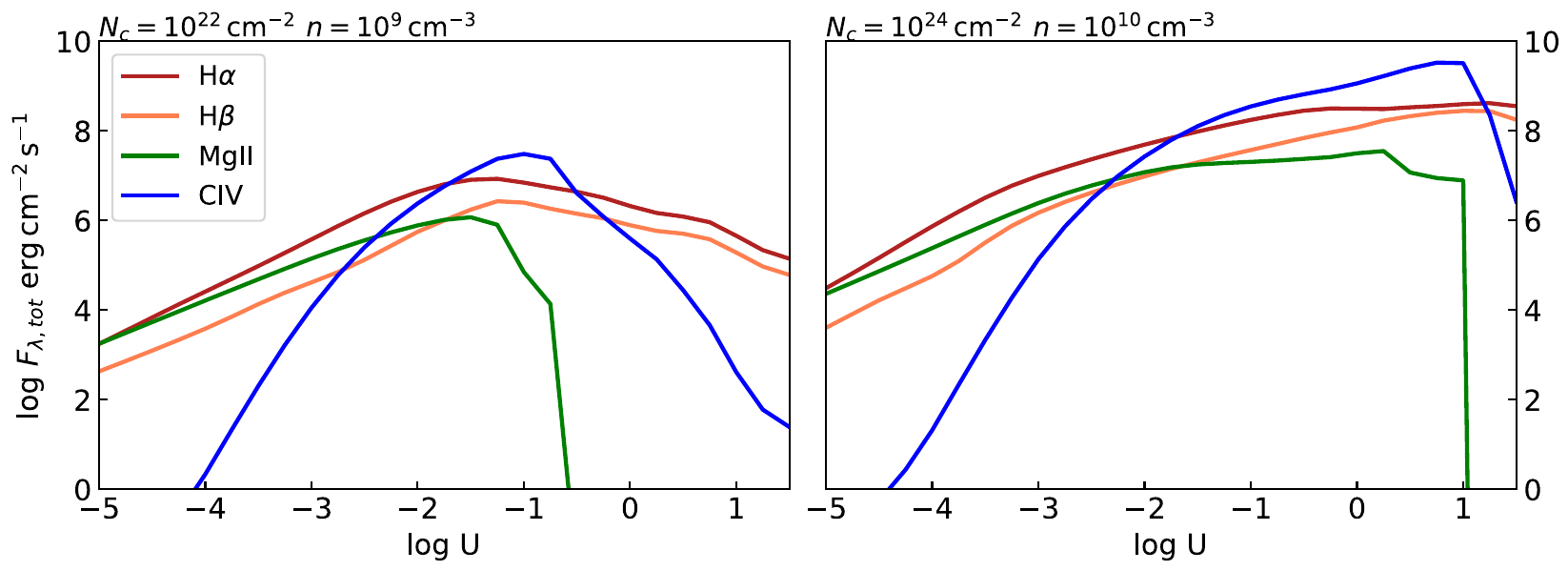}
    \caption{The total line flux emitted per unit area of the cloud surface, for the same parameters as Figure~\ref{fig:FRvU}.}
    \label{fig:FvU}
\end{figure}  

%The radiation pressure confinement model, $P_{rad}=P_{gas}$, assumes clouds are confined by a hot, inter-cloud medium, dominated by Compton heating and cooling.  However, densities based on the Compton temperature implied an optically thick BLR \citep[e.g.,][]{Mathews1987WhatNuclei,Krolik1981Two-phaseRegions}.  

Early theoretical studies of photoionized gas in the BLR suggested that the line-emitting clouds exist in thermal equilibrium with a hot, inter-cloud medium (HICM) at the Compton temperature \citep[e.g.,][]{Krolik1981Two-phaseRegions} and are confined by the pressure of the hot gas. However, at the required densities, the clouds would be unstable due to drag and the BLR as a whole would be optically thick to photoelectric absorption \citep[e.g.,][]{Mathews1987WhatNuclei}.  How the BLR clouds are confined is currently unknown; radiation pressure and magnetic fields \citep[e.g.,][]{Netzer2020TestingMapping,Baskin2018DustNuclei} are two of the most common theories. Nevertheless, it is reasonable to assume that an optically thin HICM exists, which will cause electron scattering of emission line photons traveling towards the observer.  Observational evidence for an electron scattering medium was reported by \citet{Laor2006Evidence4395}, who modeled the broad H$\alpha$ line profile of NGC 4395 and estimated an electron scattering optical depth $\tau_{e^-}\approx0.34$.  Here we assume the HICM has a uniform electron density of $n_{e^-}(r)=10^6\,$cm$^{-3}$, which is similar to estimates based on the Compton temperature ($10^{7-8}\,$K) while maintaining the condition of an optically thin environment \citep{Reynolds1995WarmNuclei,Goncalves2001EvolutionLines,Laor2006Evidence4395, Netzer2008IonizedNuclei}.  The optical depth is then $\tau_{e^-}=l\sigma_Tn_{e^-}$, where $l$ is the path length along the LOS from an individual cloud to the observer and $\sigma_T$ is the Thomson scattering cross-section. For a BLR with an outer boundary set by the dust sublimation radius at 0.4\,pc, $0<\tau_{e^-}\lesssim1.7$.  Although it is optically thick for clouds on the far side, for the majority of BLR the HICM is optically thin. Taking into account optically thin election ($e^-$) scattering, the observed luminosity of a cloud in a given emission line is 

\begin{equation}\label{eq:lum}
    L_{cl,\lambda}(\tau)=4\pi R_{cl}^2F_\lambda(n(r),N_c(r),U(r,\tau))e^{-\tau_{e^-}}.
\end{equation}
 
%Since we are using \textsc{Cloudy}'s intensity case, the luminosity of the SED will scale at each grid value of the ionization parameter, therefore $L_{AGN}$ is not necessary to create the photoionization model grids.  However, the flux values in a set of grids will depend on the shape of the SED.   

\subsubsection{Generalized Radial Motion}

In the models presented here, we assume that only the SMBH's gravity and radiation pressure contribute to the radial acceleration.  Since the origin of the BLR gas clouds is unknown, we generalize the prescription of \citetalias{Rosborough2024ModelingNuclei} to allow for scenarios where, respectively, the clouds form in a wind that is blown off the accretion disk, or originate from and are infalling from the dusty torus.  Integrating Equation~\ref{eq:eom}, the general equation for radial velocity is

\begin{equation}\label{eq:vel:rad}
   v_{rad}^2(r) = 
         \frac{2GM_\bullet}{r}\left[\frac{\Gamma F_M}{2s/3+1}\left(\left(\frac{r}{R_{in}}\right)^{2s/3+1}-1\right)+1+\frac{r}{R_o}\left(\frac{v_o(R_o)}{v_{esc}(R_{in})}-1\right) \right] 
\end{equation}      

where $v_{esc}(R_{in})$ is the escape velocity and $v_o(R_o)$ is a free parameter that determines the initial velocity of the clouds at their point of origin, $R_o$.  For an inflow, $R_o=R_d$, $v_o=v_{Kep}(R_d)$ and $v_{rad}$ is negative.  If the BLR is created by a wind from the accretion disk \citep[e.g.,][]{Elvis2000AQuasars}, the gas clouds at $R_{in}$ would have a velocity comparable to the escape velocity from the disk.  Therefore, we explore BLR models with outflows by setting $R_o=R_{in}$ and $v_o=v_{esc}(R_{in})$.  If $v_o < v_{esc}(R_{in})$, it is possible for the clouds to decelerate outward to a stalling point and then fall back towards $R_{in}$.  When \textsc{Belmac} encounters this `failed wind' condition, the clouds do not reach distances beyond the stalling point.  In this case, we assume half the clouds within the stalling radius are inflowing, and therefore the direction of their velocity vectors is reversed.  Similarly, for clouds originating at $R_d$ and infalling, the radiation pressure may be high enough to stop and reverse the direction of the inflowing clouds.  In this case as well, therefore, clouds do not exist past the stalling point.   

As mentioned in Section~\ref{sec:meth:pre}, $\Gamma F_M\propto\alpha_f$ and therefore $\alpha_f$, which depends on $U(r,\tau)$ and $N_c(r)$, influences the acceleration.  Using the incident and transmitted continua computed by \textsc{Cloudy}, we calculate the fraction of incident flux absorbed by a cloud, $\alpha_f(U(t^\prime),N_c)$, and include this fraction in the grids of photoionization models. 

\section{Results} \label{sec:results}

%PAPER 1: BELMAC can use any input light curve, but in order to obtain a numerical approximation of the transfer function, the input light curve is a single square-wave pulse to represent a $\delta$-function. We refer to the approximate transfer function as the response function \citep{Almeyda2017ModelingIllumination}.  For the duration of the pulse, $\sim4$ days, the ionizing luminosity increases by a factor of 2.  The light-front corresponding to the light pulse propagates outward from the center of the system with time and the total luminosity of the BLR is computed by integrating over the cloud ensemble at each observer time step, taking into account light travel delays.

%Here we present the time-averaged line profile ($L_\lambda(v^{||})$) and the velocity-integrated reverberation response ($\L_\lambda(\tau)$; Equation~\ref{eq:RM}).  % to a single square-wave pulse from the accretion disk.  
%\textbf{Although any input light curve can be used, the input light curve is a single square-wave pulse to represent a $\delta$-function.}  In this case, $\L_\lambda(\tau)$ is a numerical approximation of the velocity-integrated transfer function,

Here we present the time-averaged line profile ($L_\lambda(v^{||})$) and the velocity-integrated reverberation response %($\L_\lambda(\tau)$; Equation~\ref{eq:RM}) 
to an input light curve consisting of single square-wave pulse, representing a numerical approximation to a $\delta$-function. As in \citetalias{Rosborough2024ModelingNuclei}, we define the line response as,

\begin{equation}\label{eq:1drf}
L_{\lambda,\,norm}(\tau) = \frac{L_\lambda(\tau) - L_\lambda(0)}{L_{\lambda,\,max} - L_\lambda(0)}
\end{equation}

which we will refer to as the 1D response function (1DRF).  %\textbf{The light-front of the pulse propagates from the center of the BLR.  Taking into account light travel delays, the response time and the total luminosity of the BLR are computed by integrating over the cloud ensemble at each observer time step.}  The model parameters we investigate here are listed in Table~\ref{tab:para}. 
Since we will present results for H$\alpha$, H$\beta$, MgII, and CIV, we do not show the full velocity-resolved response, $L_{\lambda,norm}(\tau,v^{||})$.  However, for some models, we will present the root-mean-square (RMS) profiles to show where in the velocity-space the response amplitudes are the strongest. For a total number of time steps, $\mathcal{N}$, the RMS line profile is

\begin{equation}\label{eq:Lrms}
L_{\lambda,\,RMS}(v^{||})= \frac{1}{\mathcal{N}} \sqrt{\sum_{j=1}^{\mathcal{N}} L_{\lambda,\,norm}(\tau_j,\,v^{||})^2}.
\end{equation}

%\textbf{For a total number of velocity bins, RMS $ = \sqrt{\overline{L_{\lambda,\,norm}(\tau,\,v^{||})}^2}$}.  
We assume $v^{||}=0$\,km\,s$^{-1}$ to be the systematic velocity of the system and that the narrow emission lines, although not included in the models, are at this velocity.  These BLR models include the electron scattering opacity of the HICM ($n_{e^-}=10^6$\,cm$^{-3}$), which was not included in \citetalias{Rosborough2024ModelingNuclei}. 

We present response models for three geometry and velocity field combinations: a rotating disk, a disk-like BLR with radial flows, and a biconical radial outflow.  All of the clouds in the models explored in this paper emit anisotropically, as described by Equation~\ref{eq:flux}.  %As in \citetalias{Rosborough2024ModelingNuclei}, to isolate the response amplitude of each emission line, we subtract its minimum luminosity from the 1DRF.  
Following the same analysis as \citetalias{Rosborough2024ModelingNuclei}, we will compare the normalized response-weighted delay (RWD), calculated from the 1DRF, with the luminosity-weighted radius (LWR) for each line.  In physical units, $t^\prime_{RW}=2R_{d}$RWD$/c$ and $ct^\prime_{RW}\approx R_{LW} = 2R_{d}$LWR, where $t^\prime_{RW}$ is generally assumed to represent the time delay measured from cross-correlation analysis.  

\subsection{Responses for a Rotating Disk} \label{sec:res:KDisk}   

\begin{figure}
    \centering
    \includegraphics[width=0.8\textwidth]{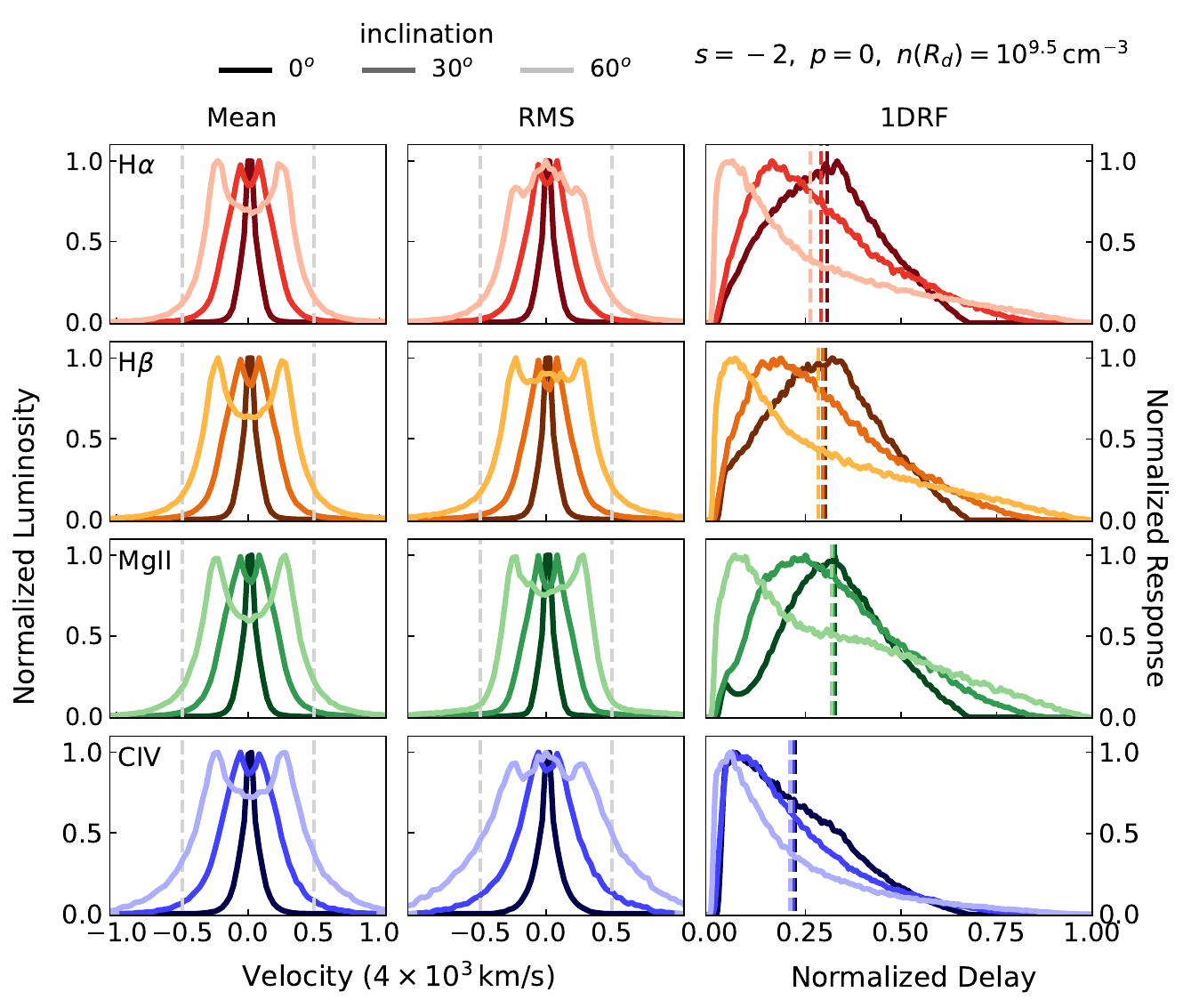}
    \caption{Rotating disk model with $s=-2$ and $p=0$.  The columns show the mean line profile (left), RMS profile (center), and 1DRF (right). From top to bottom, the rows correspond to H$\alpha$ (red), H$\beta$ (orange), MgII (green), and CIV (blue).  In each sub-panel, the shading of the colors represents the inclination of the disk, where the darkest is $i=0^\circ$.  The line profiles in each sub-panel are normalized to their respective peak luminosity.  To assist comparing the line profile shapes, the light-gray dashed lines mark $\pm2,000\,$km\,s$^{-1}$.  The 1DRFs are normalized to the maximum response amplitude for each sub-panel.  The colored dashed lines denote the RWDs of their respective 1DRF.}
    \label{fig:KDisk:s-2p0}
\end{figure}

The mean and RMS line profiles and the 1DRF for rotating disk BLR models with constant ionization parameter ($s=-2$) and a centrally concentrated cloud distribution ($p=0$) are shown in Figure~\ref{fig:KDisk:s-2p0}.  The rows of sub-panels show the results  for each of the 4 selected BELs at 3 inclinations,  $i=0\degr, 30\degr$ and $60\degr$. The RWD for each inclination is also shown in the 1DRF plots (colored dashed lines).  The line profiles of all BELs have symmetric double peaks when $i>0^\circ$, where the blueshifted and redshifted peaks result from the approaching and receding sides of the disk, respectively.  The effects of ACE and Thomson scattering are axially symmetric; therefore, for all emission lines the blue- and red-shifted peaks are identical in shape and amplitude.  As the inclination decreases, the separation between the peaks reduces and the profile becomes a single sharp peak when $i=0^\circ$.  Additionally, the wings of the profiles become less broad as the inclination decreases.  The wings are similar in width for the two Balmer lines and MgII, but slightly broader in CIV.  The RMS profiles are similar in shape, but CIV has noticeably broader wings compared to the low ionization BELs.  The emission in the wings is produced by clouds that are closest to the inner radius, with the fastest velocities and also the highest ionization parameters.  Therefore, since CIV is an HIL, its response in the wings of the profile is stronger compared to the wing responses of the LILs.  For the same reasons, the MgII RMS profiles are narrower than the Balmer RMS profiles.   

%\textcolor{teal}{In the models presented in Figure~\ref{fig:KDisk:s-2p0} the clouds are concentrated toward the center of the BLR.  Instead, as the inclination decreases, the response peaks at longer delays, showing behavior more similar to the models with more uniformly distributed clouds ($p >0$) in \citetalias{Rosborough2024ModelingNuclei}, (e.g., their Figure 16).  In the simple reprocessing models of \citetalias{Rosborough2024ModelingNuclei}, in which line emission is computed using hydrogen recombination theory, the corresponding cloud distribution ($p=0$) resulted in the 1DRFs peaking at short time delays for all inclinations (see their Figure 16\footnote{Gas densities in \citetalias{Rosborough2024ModelingNuclei} are lower by 0.5 dex and therefore the model setup here is not exactly the same.}).  These differences with inclination are most noticeable for the LILs.  For CIV, the delay at which the 1DRF peaks increases by only a small amount as the inclination changes.  This behavior can be explained by the dependence of the line flux produced by a cloud on the ionization parameter.  The models in Figure~\ref{fig:KDisk:s-2p0} have $\log\,U(r,\,\tau=0)\approx-2$, $n(R_d)\approx10^{10}\,$cm$^{-3}$, and $N_c(R_d)\approx10^{23}\,$cm$^{-2}$.  Therefore, the clouds near $R_d$ are similar to the single cloud models shown in the right panel of Figure \ref{fig:FvU}.} 

In these models the clouds are concentrated toward the center of the BLR. For the Balmer lines and MgII, the response peaks at longer delays for lower inclinations. In contrast, in the corresponding simple reprocessing models of \citetalias{Rosborough2024ModelingNuclei} the 1DRFs peak at short time delays for all inclinations (see their Figure 16\footnote{Gas densities in \citetalias{Rosborough2024ModelingNuclei} are lower by 0.5 dex and therefore the model setup here is not exactly the same.}). In fact, the behavior of the LILs is generally similar to the \citetalias{Rosborough2024ModelingNuclei} models that have more uniformly distributed clouds ($p>0$). %Although these 1DRFs are noticeably different compared to their respective models in \citetalias{Rosborough2024ModelingNuclei}, the RWDs are similar. For the LIL 1DRFs here, RWD $\sim0.3$ and for the corresponding model in \citetalias{Rosborough2024ModelingNuclei} RWD $\sim0.25$. 
These differences in response behavior and RWD are attributed to the use of photoionization models. 

On the other hand, the variations in the 1DRF with inclination are much smaller for CIV, and closer to the corresponding ($p=0$) \citetalias{Rosborough2024ModelingNuclei} models. As discussed in \citetalias{Rosborough2024ModelingNuclei}, ACE suppresses the emission from clouds nearest to the observer, and therefore the 1DRF is more extended with time and peaks at longer delays.  However, since CIV is largely an optically thin emission line, the effect of ACE on the 1DRF is much smaller compared to the optically thick LILs.  For all BELs, the 1DRFs decrease rapidly for $\tau\gtrsim 0.5$ as the optical depth through the HICM increases for clouds responding progressively further from the observer.  The result is that the CIV response is dominated by emission close to the center of the BLR, where inclination changes have less of an impact.  Therefore, the CIV 1DRFs are similar for all 3 inclinations in Figure~\ref{fig:KDisk:s-2p0}. 

\subsubsection{Negative Responses}\label{res:KDisk:invrs}

\begin{figure}[h!]
\centering
\includegraphics[width=0.8\columnwidth]{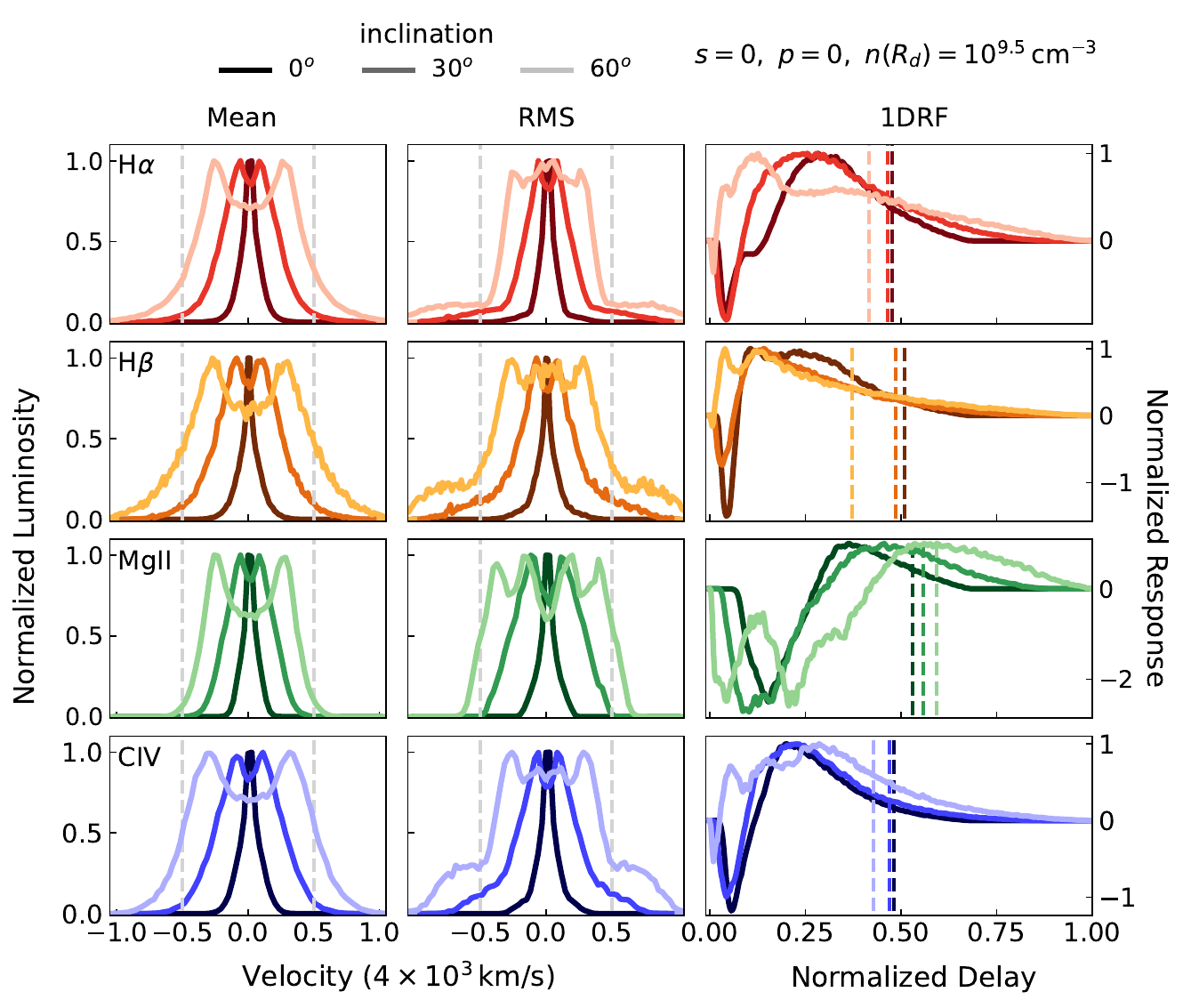}
\caption{The same as Figure~\ref{fig:KDisk:s-2p0}, but for BLR model parameters $s=0$ and $p=0$.}  
\label{fig:KDisk:s0p0}
\end{figure}

Figure~\ref{fig:KDisk:s0p0} presents models with $n(R_d)=10^{9.5}\,$cm$^{-3}$, $p=0$, and $s=0$.  Although the mean and RMS profiles are similar to those of the $s=-2$ case, it is notable that the 1DRFs initially have a negative response.  In these instances, a negative response indicates that the line emission of the BLR decreases locally as the ionizing continuum luminosity and hence $U(r,\,\tau)$ increases.
%This behavior has also been encountered and investigated in previous modeling work by \citet{Sparke1993DoesUp, Goad1993ResponseNuclei, OBrien1994ResponseEmission} for similar BLR model parameters.}   

\citet{Sparke1993DoesUp} has argued that reverberation mapping data of NGC\,5548 show indications that line strengths in the inner part of the BLR decrease as the continuum strengthens -- i.e., exhibit a negative response. In their extensive investigation of BEL response functions using a similar cloud population photoionizaton model, \citet{Goad1993ResponseNuclei} also encountered situations in which some lines have negative responses. In subsequent work, \citet{OBrien1994ResponseEmission} discussed the combined effects of ACE and negative responsivity on the line response functions.

Returning to our models, in this case, the gas density and the cloud column density are constant throughout the BLR with values $n(r)=10^{9.5}\,$cm$^{-3}$ and $N_c(r)=10^{22.25}\,$cm$^{-2}$, similar to the models shown in the left panels of Figures~\ref{fig:FRvU} and \ref{fig:FvU}. The ionization parameter decreases with radius as $U(r,\,\tau=0)\propto r^{-2}$. The response for all four BELs is initially negative, meaning that the line luminosities initially decrease in response to the continuum pulse, then becomes positive as the delay increases.  This negative response is due to the high ionization parameter in the inner region of the BLR, which prior to the onset of the pulse is $\log\,U(R_{in},\,\tau=0)=0.56$. As shown in Figure~\ref{fig:FvU}, the line fluxes decrease as $U$ increases for $\log\,U\gtrsim-1$.  Furthermore, the CIV and the Balmer lines also become optically thin ($F_{in}/F_{tot}=0.5$, Figure~\ref{fig:FRvU}), which nullifies the ACE effect. As a result, there is a sharp drop in the 1DRF, which reaches a minimum at a short delay of $\tau=1/Y_{BLR}$.  As the inclination increases, more clouds at greater radial distances whose line fluxes increase with $U$ are intercepted by the observer's LOS and therefore respond at earlier delay times.  Their contribution decreases the amplitude of the negative response.  In the case of MgII, when $i=0^\circ$ the onset of the response is delayed and the negative response persists for longer time delays than for the other BELs. The reason is that in the inner regions of the BLR, where $\log\,U(r,\,\tau)\gtrsim -0.5$, the ionization parameter is too high for MgII emission to be produced at all (Figure~\ref{fig:FvU}).  As the response front propagates through the BLR, $U(r,\,\tau)$ decreases and the response begins to behave ``normally", i.e., positively correlated with the continuum pulse.         

The negative response portion of the 1DRFs also has the effect of increasing the RWD. Therefore, the RWDs for these models are nearly twice as long as the RWDs for the constant $U$ ($s=-2$) models in Figure~\ref{fig:KDisk:s-2p0}, which do not have negative responses.  Further analysis of RWDs is presented in Section~\ref{sec:res:rwds} and discussed in Section~\ref{sec:dis:mass}.     

%The mean and RMS line profiles appear very similar to the models of Figure~\ref{fig:KDisk:s-2p0}.  

%rewrote and condensed above: The two Balmer lines have similar 1DRFs shapes for corresponding inclinations.  As discussed in \citet{Rosborough2024ModelingNuclei}, the response begins at increasing $\tau$'s as $i$ decreases for disk BLR models due to a lack of clouds at the BLR's poles.  This added time delay is mainly a geometry consequence, however the start time of the response is indirectly related to wavelength.  The Balmer and CIV lines have equal delayed start times for their corresponding $i$ model values.  At low inclinations, MgII has longer response start times than the other three lines.  Since MgII is a LIL, $U(\tau,r)$ is too high for MgII emission.  The 1DRFs for low $i$ start when $\log\,U(\tau,r)<-0.5$, given the $n(r)$ and $N_c(r)$ values.  % conversely for CIV, but maybe save that for later as this case is not shown here.
%For the Balmer lines, as $i$ \emph{increases}, the time when the response is most negative decreases and becomes more sharp.  For MgII, the opposite pattern occurs, as $i$ \emph{decreases}, the time when the response is most negative decreases and becomes more sharp.  Additionally, when $i=60\degr$, the 1DRF has 2 negative response dips.  % this is due to elf because...  

\subsection{Responses for Radial Flows in a Disk} \label{sec:res:RDisk}  

\begin{figure}
    \centering
    \includegraphics[width=0.8\textwidth]{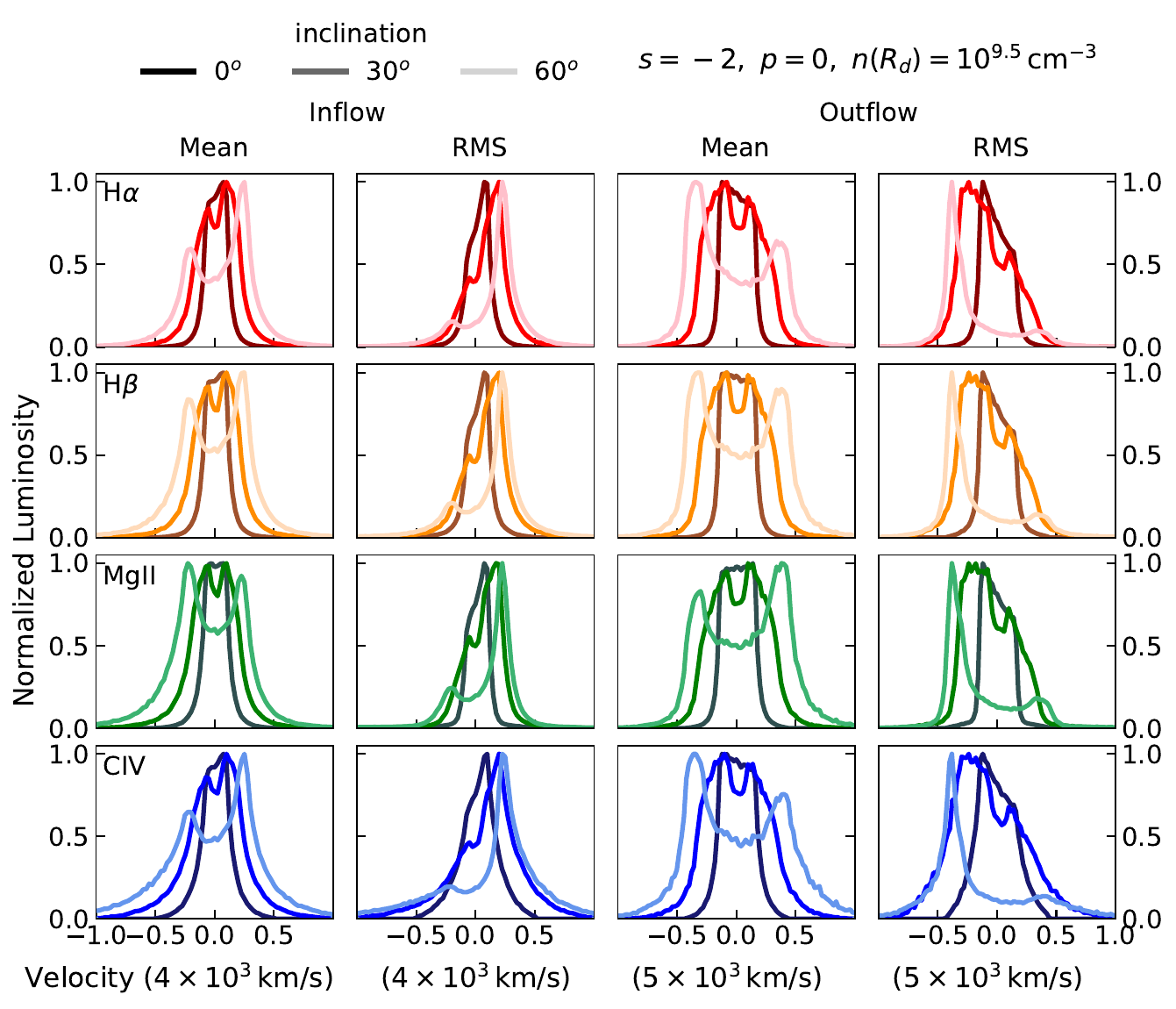}
    \caption{The mean line profiles and RMS profiles for disk-like BLRs with a radially inflowing (left two columns) and outflowing (right two columns) velocity fields. All of the models have $s=-2,~p=0$, and $\log\,n(R_d)=9.5\,$cm$^{-3}$. The line color and shade indicate the specific BEL and inclination as in Figure~\ref{fig:KDisk:s-2p0}.}
    \label{fig:RDisk:s-2p0}
\end{figure}

Here we present the responses of a disk-like BLR with outflow and inflow velocity fields for the parameters listed in Table~\ref{tab:para}. The mean and RMS line profiles for models with $s=-2,~p=0$, and radial in- and outflows are presented in Figure~\ref{fig:RDisk:s-2p0}.  These models have the same 1DRFs as the rotating disk models shown in Figure~\ref{fig:KDisk:s-2p0}, as the geometry, cloud distribution, and gas density parameters are identical. The clouds are, respectively, accelerating inward (left 2 columns) or decelerating outward (right 2 columns).  The clouds in these models have gas densities and column densities similar to those shown in the right-hand panel of Figure~\ref{fig:FvU} and have $\log\,U(r,\,\tau)\gtrsim-1.7$.  

The mean line profiles are double-peaked, similarly to those of the rotating-disk models, but in this case, the peaks show varying degrees of asymmetry.  In general, the peak/core region of the mean line profiles is red-asymmetric in the inflow case (the red peak is stronger) and blue-asymmetric for the outflow (the blue peak is stronger). MgII is the exception, exhibiting the opposite behavior. However, in all 4 lines and most noticeably at higher inclinations ($i\geq30^\circ$), the blue wing is more extended for the inflow models, while the red wing is more extended for the outflow models.
These asymmetries are due to the combination of the HICM obscuring the clouds farthest from the observer and the fact that individual clouds emit anisotropically.  
%ACE is calculated from the cloud's position in the BLR and the emission directed toward and away from the ionizing source (i.e., from the illuminated and non-illuminated surface of the cloud, respectively), as determined from the photoionization models, therefore, ACE is directly related to line opacity.  

\begin{figure}
    \centering
    \includegraphics[width=0.85\textwidth]{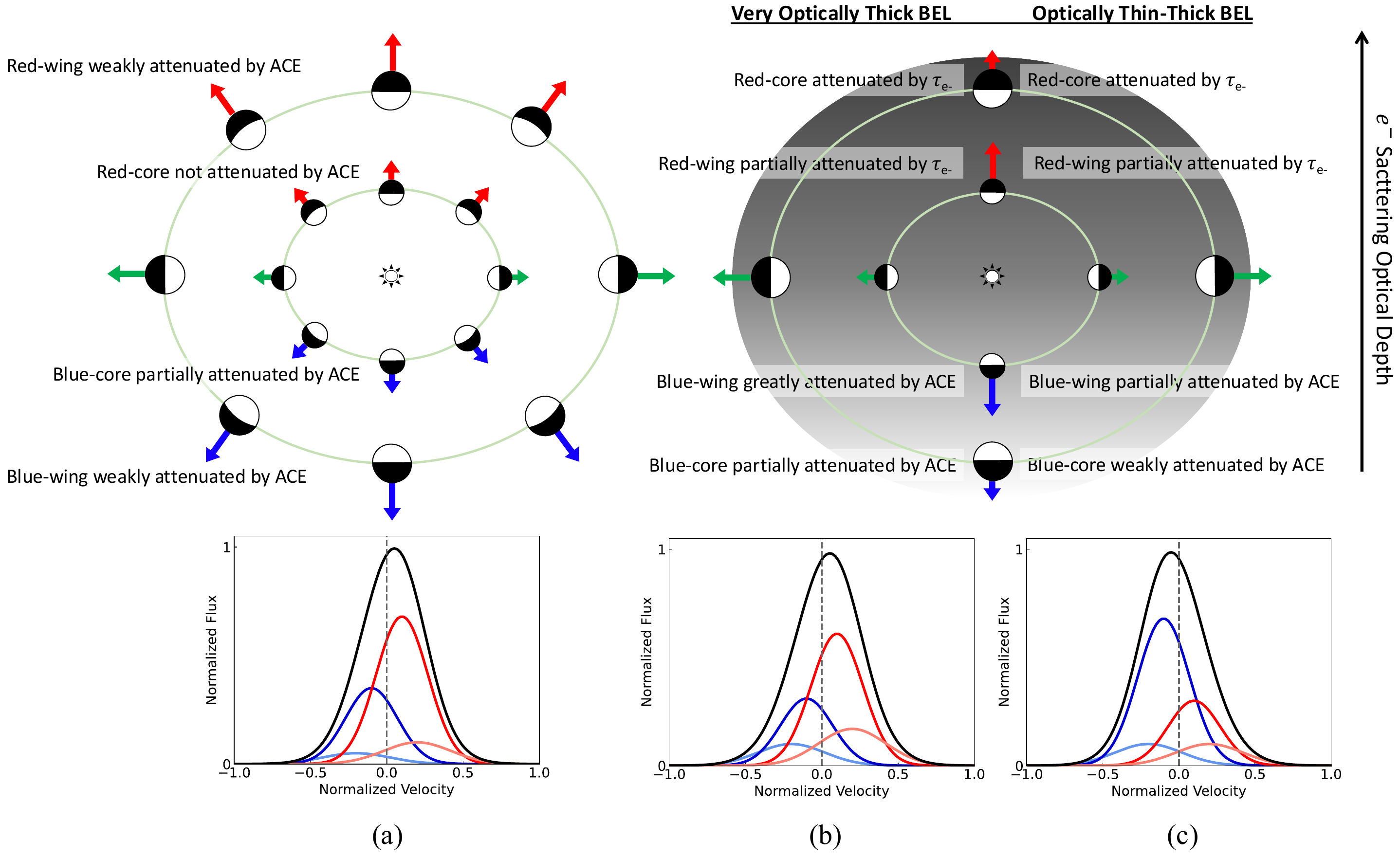}
    \caption{Schematic illustration of the 2 causes of BEL profile asymmetry in the radial in/outflow models: ACE and Thomson scattering in the hot, inter-cloud medium. In both diagrams, the continuum source is located at the center and the observer is located below the figure. The representative spherical clouds located around the rings are emitting anisotropically, with the non-illuminated face shaded black. Unless gas density is constant, the cloud size increases with distance.  \emph{Left:} The effects of ACE without Thomson scattering in the HICM.  The clouds are accelerating outwards, with the length of the arrows indicating the velocity magnitude and the color the direction.  As shown in panel (a), ACE suppresses the blue side of the line profile, which is therefore red-asymmetric, but has a blue-wing.  \emph{Right:} The combined effects of ACE and Thomson scattering on the line profile shape.  The gray fade represents the gradient of the electron scattering optical depth with respect to the observer, where darker shades indicate a greater optical depth.  The clouds are decelerating outwards. Panel (b) illustrates the resulting profile shape for a very optically thick emission line (such as MgII), which is more affected by ACE than the HICM and therefore is red-asymmetric and has a red-wing. Panel (c) illustrates the resulting profile shape for a moderately thick or thin emission line (such the Balmer lines and CIV, respectively).  The HICM opacity affects the line emission more than ACE and the profile is therefore blue-asymmetric, but has a red-wing.  The two sources of opacity primarily affect opposite sides of the BLR, and thus may compete in modifying the shape of the line profile.}
    \label{fig:RDisk:asy}
\end{figure}
% add normalized LP, vel. normalized, more description of LP plots to caption, add blue/red shift indication.

To analyze the origin of the line profile asymmetries, we first consider a situation where the BLR has no HICM, but with the clouds emitting anisotropically, as depicted in the left-hand diagram in Figure~\ref{fig:RDisk:asy}.  When clouds are outflowing, the blueshifted emission is suppressed, since the observer primarily views the non-illuminated face of the near-side clouds.  Instead, most of the emission is observed from the receding clouds on the far side of the BLR and the line profile core shape is therefore red-asymmetric.  Conversely, for an inflowing velocity field, the line profile core is blue-asymmetric.  

The asymmetry of the profile wings depends on whether the radial flow is accelerating or decelerating.  Clouds at or near the inner radius are most affected by ACE, since they have the smallest radii (for $s=-2$, cloud radius $R_{cl}\propto r^{4/3}$).  Thus, the area of the illuminated crescent is smaller for clouds located towards the center of the BLR, than it is for the outermost clouds near $R_d$.  For an accelerating outflow, as shown by the arrows in Figure~\ref{fig:RDisk:asy} (a), the fastest clouds are at the outer radius, and the blue-wing is only weakly suppressed compared to the blue side of the profile's core, which is produced by the slower inner clouds.  The result is a red-asymmetric profile with a red-wing \emph{and} a weaker blue-wing (see lower panel in Figure~\ref{fig:RDisk:asy}(a)).  How severely the blue wing and core are each suppressed depends on the cloud distribution and velocity field.  When the clouds are concentrated near the center, the blue-core is more strongly suppressed compared to when they are distributed uniformly.  This will affect the profile's degree of red-asymmetry.  If the outflow is decelerating, the profile would still be red-asymmetric, but the blue-wing would be strongly suppressed, whereas the blue-core would be only partially suppressed. In the models of Figure~\ref{fig:RDisk:s-2p0}, the BLR has a centrally concentrated cloud distribution. Therefore, the accelerating inflow models have line profiles with extended blue-wings and the decelerating outflow models have line profiles with extended red-wings.        

The right-hand diagram of Figure~\ref{fig:RDisk:asy} illustrates the combined effect of ACE and scattering by a HICM for (b) highly optically thick and (c) optically thin to moderately thick BELs. Additionally, in this case, we consider a decelerating outflow, to demonstrate the differences in the shapes of the core and wings in comparison to the accelerating outflow shown in the left-hand diagram. As the optical depth due to electron scattering in the HICM is proportional to the LOS path length, the emission from the outer clouds on the far-side of the BLR suffers correspondingly greater attenuation.  In contrast, ACE most strongly affects the innermost clouds on the near side. ACE and the HICM, therefore, have competing effects that determine the asymmetry of the core and wings of the line profiles. The two opacity effects combine to explain the asymmetries of the BEL shapes shown in Figure~\ref{fig:RDisk:s-2p0}. For optically thin to moderately thick BELs the effect of line opacity (ACE) is comparatively weak compared to scattering in the HICM (Figure~\ref{fig:RDisk:asy}(c)) and thus the line profile has a blue-asymmetric core, as seen in the Balmer and CIV profiles in Figure~\ref{fig:RDisk:s-2p0}.  However, for very optically thick BELs, the line opacity has a greater influence on the profile asymmetry than the HICM (Figure~\ref{fig:RDisk:asy}(b)).  Therefore, despite the outflowing velocity field, the line profile has a red-asymmetric core, as does MgII in Figure~\ref{fig:RDisk:s-2p0}. If the velocity field was inflowing, rather than the outflow depicted in Figure~\ref{fig:RDisk:asy}, the asymmetries of these two types of BELs would be reversed.  

In summary, with a uniform distribution of the HICM ($n_e(r)=$ constant in the current version of \textsc{Belmac}), the line profile asymmetry in the core and wings depends on the line opacity, the cloud distribution, and velocity field. We also expect these results to be illustrative for cases in which the density distribution of the HICM is not uniform, although the relative importance of the HICM and ACE will then also depend on $r$.
% radius.

Although the asymmetries of the mean line profiles vary between BELs and with flow direction, the RMS profiles are always blue-asymmetric for an outflow and red-asymmetric for an inflow.  This is because the response begins in the blue-wing for a decelerating outflow and red-wing for an accelerating inflow.  The asymmetry is apparent for all inclinations and, even for $i=0^\circ$, is quite strong.  Section~\ref{sec:dis:asym} provides a discussion of the asymmetries of the mean and RMS profile shapes with respect to the observed BELs.

\subsection{Response Weighted Delays and Luminosity Weighted Radius}\label{sec:res:rwds} 

 \begin{figure}[h!]
    \centering
    \includegraphics[width=0.75\textwidth]{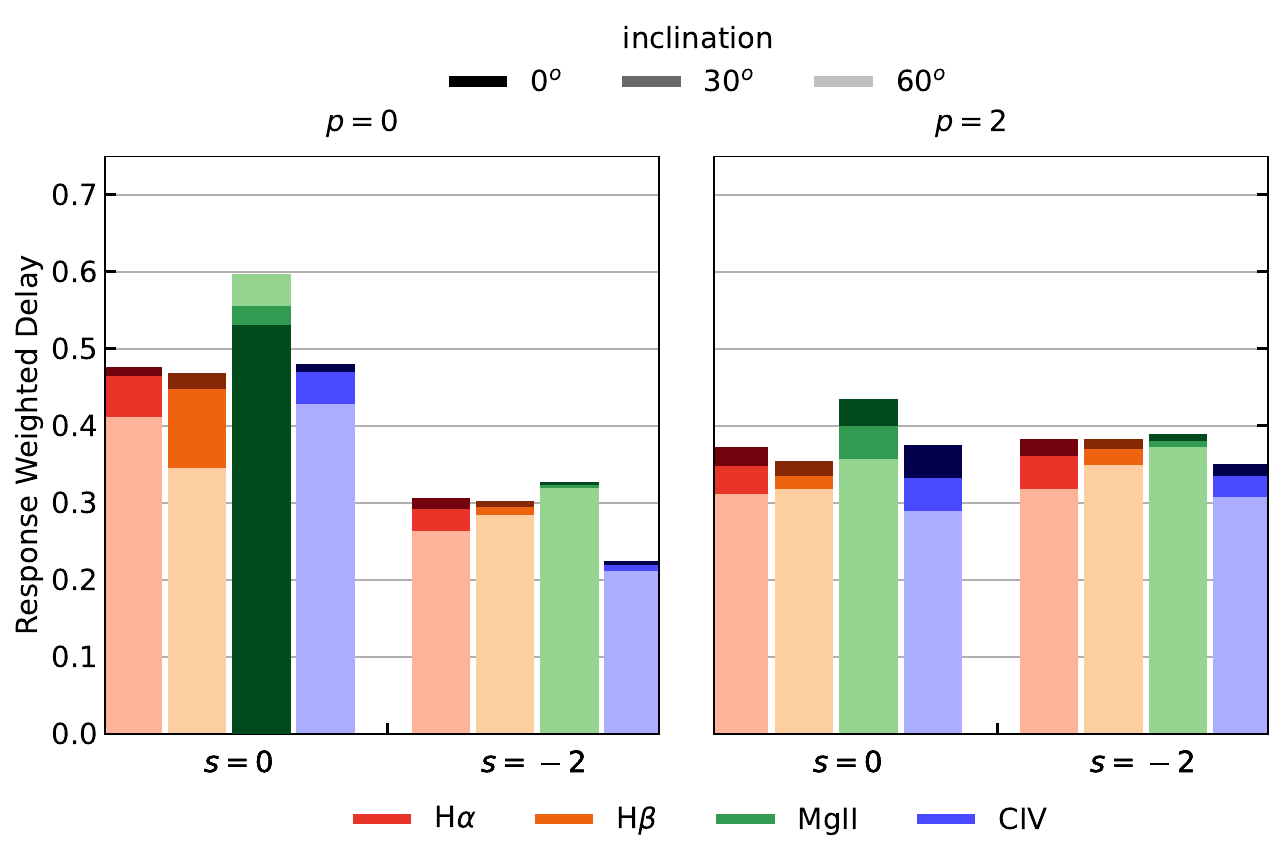}
    \caption{The RWDs for disk models with $p=0$ (left) and $p=2$ (right).  The subplots are divided in 2 groups of models with $s=0$ and $-2$, with the values for the 4 BELs further grouped by inclination and where the top of each bar is the RWD value.  The color and shade of each bar denotes the emission line and inclination, the same as e.g., Figure~\ref{fig:RDisk:s-2p0}.
    }
    \label{fig:Disk:lags}
\end{figure}

\begin{figure}[h!]
    \centering
    \includegraphics[width=0.8\textwidth]{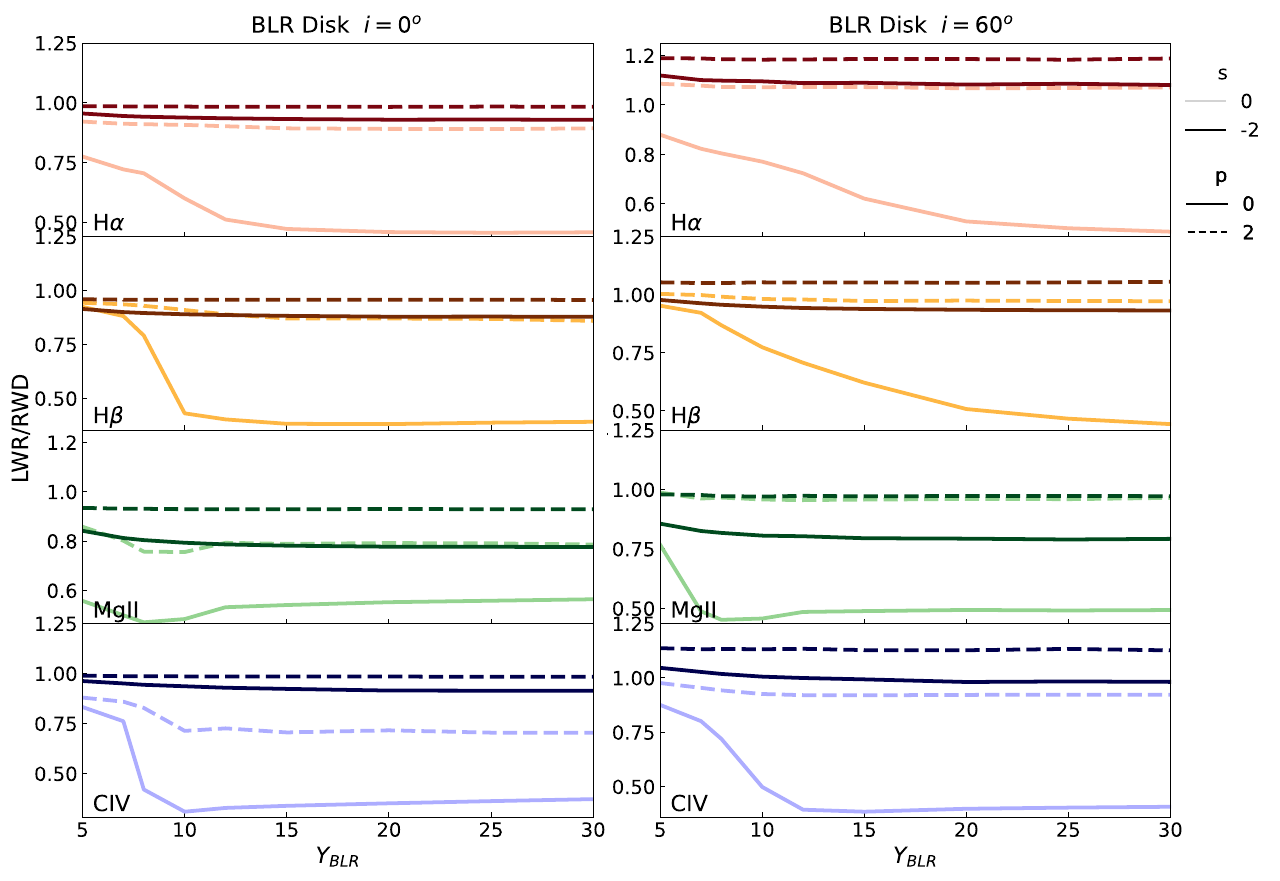}
    \caption{The ratio of LWR to RWD for disk models with varying $Y_{BLR}$ for, from top to bottom, H$\alpha$ (red), H$\beta$ (yellow), MgII (green), and CIV (blue). The light and dark shaded lines denote models with $s=0$ and $-2$, respectively.  The solid and dashed lines denote models with $p=0$ and 2, respectively. The left- and right-hand columns show models for inclinations $i=0^\circ$ and $60^\circ$, respectively.}
    \label{fig:Disk:rwd/lwr}
\end{figure}

Figure~\ref{fig:Disk:lags} presents, for each BEL, the RWDs for all the BLR disk models explored in this work. The left and right subplots show models with $p=0$ and $2$, respectively. 
In the $p=2$ case, when clouds are uniformly distributed, all 4 BELs have similar values of RWD for all combinations of $s$ and $i$, with CIV typically having slightly shorter RWD's (by $\lesssim 10\%$). In the $p=0$ case, when the clouds are more centrally concentrated, all lines have RWD's that are either longer (for $s=0$), or shorter (for $s=-2$), than in the corresponding $p=2$ models. Compared to the other lines, CIV has RWD's that are much shorter (by $\sim30\%$) when $s=-2$ but are comparable with the Balmer lines for $s=0$. The RWDs of MgII are consistently longer (by $\gtrsim 10\%$) than for the other BELs, especially in the $p=0$, $s=0$ models.  In particular, the models with $s=0$ and $p=0$ produce the longest RWDs in all lines, for a given inclination. These are the models that produce negative responses at early time delays, leading to longer RWD's. This effect also accounts for the slightly longer RWD's of MgII in the $s=0$, $p=2$ models, since this line exhibits a weak negative response for these parameters.
%In the $p=2$ case, H$\alpha$ and H$\beta$ have roughly equal RWD values, for combinations of $s$ and $i$.  Compared to the other lines, MgII has slightly longer RWDs and CIV has the shortest (by $\lesssim 10\%$). When clouds are uniformly distributed ($p=2$) \{and did not respond inversely (not the $p=0$ and $s=0$ case)}, the RWDs for all BELs are slightly longer compared to when clouds are concentrated toward the center ($p=0$).  Notably, CIV has much shorter delays, $\sim40\%$ shorter compared to when $p=2$ and $\sim30\%$ shorter compared to the other BELs when $p=0$.  Additionally, when clouds are centrally concentrated, there is only a small difference in RWDs between models of different inclination.  When $s=0$, the RWDs for the Balmer lines are also approximately equal for the same $p$.  However, the delays for CIV are also similar to the Balmer lines, rather than shorter.  The RWDs for MgII are consistently longer (by $\gtrsim 10\%$) than for the other BELs.  

The LWR/RWD ratio is shown as a function of BLR size ($Y_{BLR}$) in Figure~\ref{fig:Disk:rwd/lwr} for the disk models listed in Table~\ref{tab:para}, with inclinations $0^\circ$ and $60^\circ$. As discussed in \citetalias{Rosborough2024ModelingNuclei}, we expect LWR/RWD $=1$ if the reverberation lag measured from cross-correlation analysis accurately yields the luminosity weighted-radius of the BLR. In general, the LWR/RWD ratio decreases with increasing $Y_{BLR}$, although only slightly for most combinations of $s$ and $p$.  The ratio also increases slightly with inclination.  With the exception of the $s=0,~p=0$ case, the LWR/RWD ratios for the Balmer lines are close to 1, regardless of inclination or $Y_{BLR}$, while the ratios for MgII and CIV show more spread with the $s$ and $p$ combinations, for example, for CIV, LWR/RWD\,$\approx 0.75$ for $p=2, s=0$. The models in which the lines exhibit a negative response to the continuum pulse (e.g., Figure~\ref{fig:KDisk:s0p0}) are a striking exception.   In these cases ($p=0, s=0$), LWR/RWD quickly drops to values $\lesssim 0.5$ as $Y_{BLR}$ increases for $i=0^\circ$ and less rapidly for $i=60^\circ$.  As $Y_{BLR}$ increases, the inner radius becomes smaller and more clouds are exposed to a high ionizing flux, which increases the intensity and duration of the negative response.    

\subsection{Radial Flows in a Bicone} \label{sec:res:RCone}

\begin{figure}[h!]
    \centering
    \includegraphics[width=0.9\textwidth]{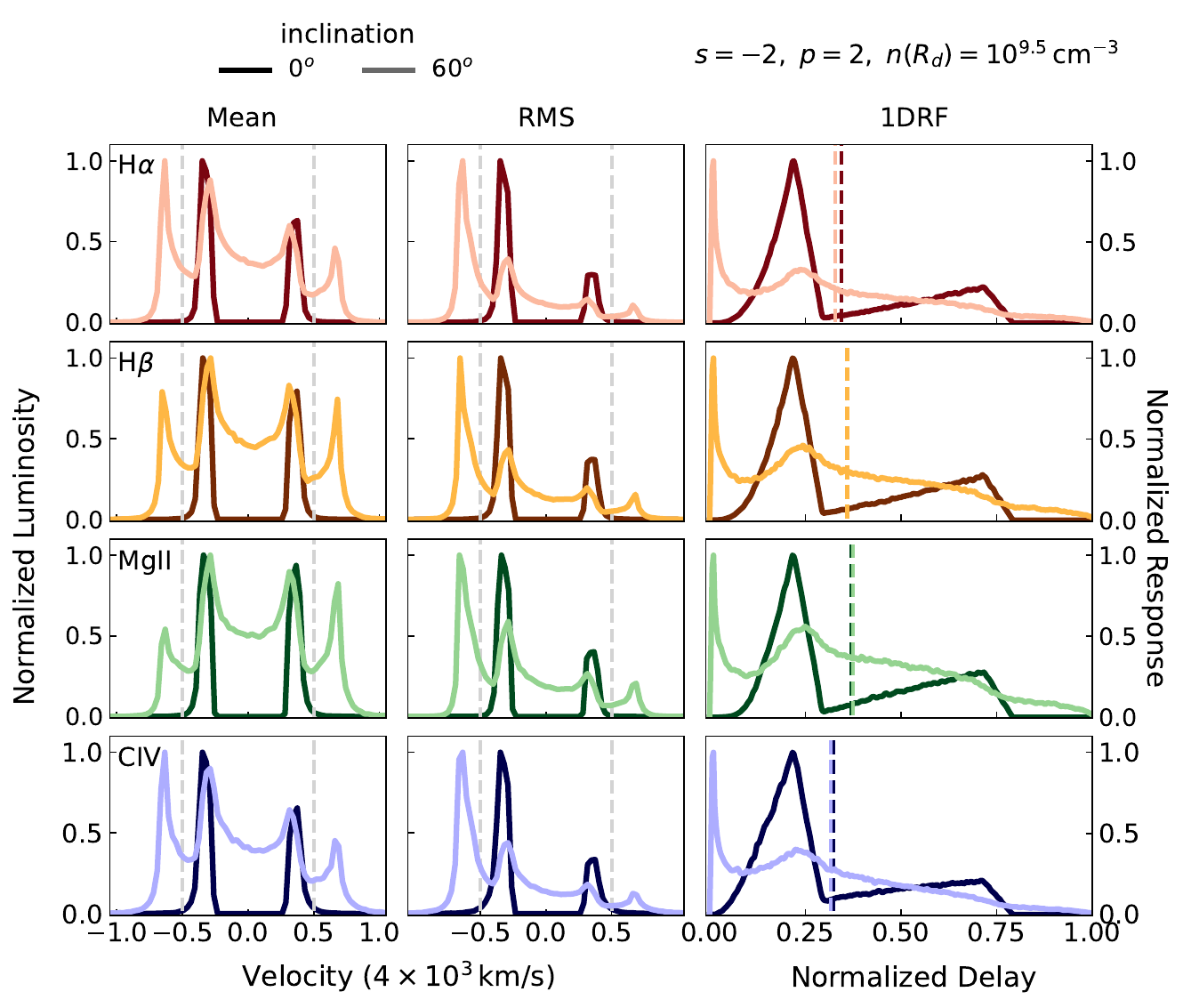}
    \caption{Biconical outflow model with $s=-2$ and $p=2$. The columns show the mean line profile (left), RMS profile (center), and 1DRF (right). From top to bottom, the rows correspond to H$\alpha$ (red), H$\beta$ (orange), MgII (green), and CIV (blue). Each plot shows models at inclinations of $i=0^o$ (darker shading) and $i=60^o$ (lighter shading). The velocity scale is in units of $3,000\,$km\,s$^{-1}$. The vertical dashed lines are as described in the caption to Fig.~\ref{fig:KDisk:s-2p0}}
    \label{fig:RCone}
\end{figure}

Here we present the responses for models in which the BLR is a biconical radial outflow defined by the parameters listed in Table~\ref{tab:para}, with $s=-2$ and $p=2$.  Figure~\ref{fig:RCone} displays the mean line profiles and the 1DRFs for $i=0^\circ$ and $60^\circ$.    

When $i=0^\circ$, the observer is looking along the symmetry axis through the (empty) funnels of the near- and side cones.  At this orientation, there are no clouds moving perpendicular to the observer's LOS. Therefore, there is no emission around $v^{||}=0$\,km\,s$^{-1}$ in the mean line profiles.  There is also a slight delay in the start of the 1DRF before the light-front reaches the inner walls of the near-sided cone.  The sharp blue- and red-shifted peaks of the line profiles are, respectively, due to the near- and far-sided outflows moving towards and away from the observer.  Likewise, the first and second peaks of the 1DRFs are the respective responses of the near- and far-sided cones.  The second peak is more extended with $\tau$ for all BELs and much weaker than the initial peak.  The extension with time delay of the second peak is due to ACE and its low amplitude is due to the increasing optical depth of the HICM with path length. 

\begin{figure}[h!]
    \centering
    \includegraphics[width=\textwidth]{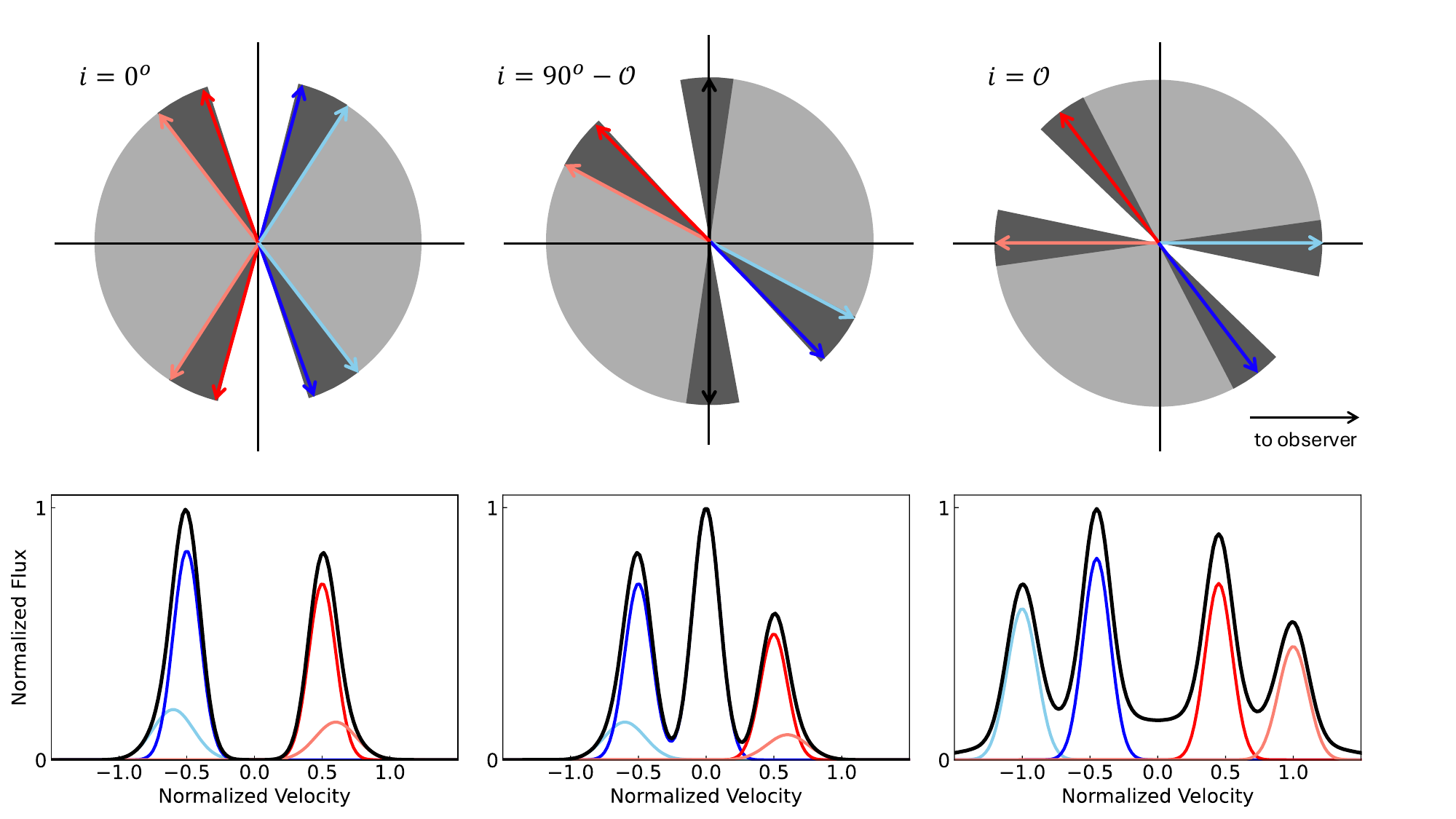}
    \caption{Creation of multiple profile peaks in biconical flows.  Cross-sectional diagrams of outflows in bicones with half-opening angles, $\mathcal{O}=60^o$, and their produced line profile for $i=0^\circ,~90^\circ-\mathcal{O}$, and $\mathcal{O}$, from left to right, respectively.  The arrows represent the velocity of the gas producing the brightest peaks in the line profiles.}
    \label{fig:RCone:geo}
\end{figure}

When $i=60\degr$, the mean line profiles exhibit 2 red-shifted and 2 blue-shifted peaks. This is a geometrical effect, which is illustrated in Figure~\ref{fig:RCone:geo}.  Since the half-opening angle of the bicone is $\mathcal{O}=60^\circ$, at $i=60^\circ$,  one side of both the near and far-sided outflows is aligned with the observer's LOS (see also Figure~\ref{fig:geo}).  These two regions produce the emission peaks at the fastest blue- and red-shifted velocity.  The clouds located in the opposite side of each cone ($\pm30^\circ$ from the LOS) produce the two peaks at about half the maximum LOS velocity.  We do not show the results for the $i=30^\circ$ model in Figure~\ref{fig:RCone} to avoid overcrowding of the plot, but the geometry is shown in Figure~\ref{fig:RCone:geo}.  In this case, one ``wall'' of each cone is perpendicular to the LOS, creating a strong central peak at $v^{||}=0$\,km\,s$^{-1}$ with asymmetric, weaker blue- and red-shifted peaks on either side.                    

In general, the mean line profiles produced by these biconical outflows are  asymmetric, with the red- and blue-shifted peaks at equal $|v^{||}|$ having different amplitudes. In particular, in H$\alpha$, H$\beta$ and CIV the blue-shifted peaks have larger amplitudes than their red-shifted counterparts. The exception is MgII, an optically thick line, which has slightly red-asymmetric outer peaks and blue-asymmetric inner peaks. These differences in the degrees of asymmetry between the outer and inner peaks arise due to the combination of the cloud distribution, velocity field, and the line opacity effects. They are explained in the same way as the wing and core asymmetries of the disk/radial flow profiles discussed in Section~\ref{sec:res:RDisk} (Figure~\ref{fig:RDisk:s-2p0}). In contrast, as the response always begins in the blue-shifted gas for an outflow, the RMS profiles for all of the emission lines are highly blue-asymmetric.

\section{Discussion} \label{sec:dis}

\subsection{Comparing Models to Observed HILs and LILs} \label{sec:dis:bels}

Numerous observations of AGN BELs, both single-epoch and time-domain spectra, have revealed commonly occurring differences between the shapes of the high and low ionization lines.  Here we compare our results with the observed differences between HILs and LILs listed in Section~\ref{sec:intro}, which we reiterate here:  

\begin{enumerate}
    \item HILs tend to be blueshifted with respect to the LILs \citep[e.g.,][]{Richards2011UNIFICATIONEMISSION,Espey1989H-alphaQuasars, Marziani1996ComparativeNuclei, Sulentic2000PhenomenologyNuclei, Baskin2005WhatNuclei}.

    The models show no median velocity shift between the HILs and LILs; the median velocities of all mean profiles fall within the central velocity bin ($v^{||}=0$\,km\,s$^{-1}$, $\Delta v^{||}=100\,$km\,s$^{-1}$).  If, on the other hand, we measure velocity shifts from the peaks of the line profiles, the CIV line is blueshifted in the outflow models by up to $v^{||}\approx2,000\,$km\,s$^{-1}$.  In contrast, for some outflow models (e.g., Figure~\ref{fig:RDisk:s-2p0}), the MgII peak is redshifted by $v^{||}\approx1,000\,$km\,s$^{-1}$. However, the peaks of the other LILs, H$\alpha$ and H$\beta$, are blueshifted to about the same extent as CIV.  In all of the \textsc{Belmac} models presented in Section~\ref{sec:results}, which include just a single velocity field component, there are no instances in which CIV is blueshifted relative to all of the LILs.         
    
    \item HILs tend to be broader than LILs \citep[e.g.,][]{Sulentic2000PhenomenologyNuclei}.
    
    The single-zone models with a single velocity component (Keplerian or radial motion) produce BELs with comparable widths; there are no large differences in full-width, half-maxima (FWHM) between HILs and LILs.  In cases where the clouds are concentrated towards the center of the BLR, the wings of CIV are somewhat broader compared to those of MgII.  However, when the gas density is relatively low, as it is for the models presented in Figure~\ref{fig:KDisk:s0p0}, the Balmer lines also have wings that are slightly broader than those of MgII and comparable to CIV.    
    
    \item LILs tend to be symmetric while HILs tend to be blue-asymmetric \citep[e.g.,][]{Baskin2005WhatNuclei,Richards2002BroadQuasars}.
    
    The profile shapes produced by disk models with circular Keplerian orbits presented in Section~\ref{sec:res:KDisk} are always symmetric. Asymmetric profile shapes result only from models including radial flows.  None of the models with radial flows presented in Section~\ref{sec:res:RDisk} and \ref{sec:res:RCone} have perfectly symmetric line profiles. However, Figure~\ref{fig:RDisk:s-2p0} demonstrates a case of radial inflow in a disk at $i=30^\circ$ that produces nearly symmetric MgII profile shapes, while all other lines are asymmetric, with CIV showing the strongest asymmetry.  The opacity effects associated with ACE and the HICM counteract one another; this suggests that symmetric LILs and asymmetric HILs could arise from a radial velocity field if the distribution of the HICM is non-uniform (see Section~\ref{sec:dis:asym}). 
    
    \item HILs tend to respond on shorter time delays than LILs. The wings of LIL profiles tend to vary on shorter time-scales compared to the core \citep[e.g.,][]{Horne2021Space5548, Bentz2010THELINES}.
    
    Figure~\ref{fig:Disk:lags} shows that CIV tends to have the shortest RWD compared to LILs, regardless of inclination or how the clouds are distributed.  However, if the gas density is constant throughout the BLR ($s=0$), CIV can have RWDs similar to the other BELs.  As explained in \citetalias{Rosborough2024ModelingNuclei} the response of a rotating disk model begins in the core of the line profile, then rapidly and symmetrically propagates to the full width of the wings, then gradually moves inwards to the core of the profile.  As shown by the RMS profiles of Figures~\ref{fig:KDisk:s-2p0} and \ref{fig:KDisk:s0p0}, this symmetrical pattern occurs for all BELs in models with circular cloud orbits.  For radial flow models, the response begins in either the red or blue wing depending on the direction of the flow, then gradually moves across to the other wing. Therefore, the rotating disk models are consistent with reverberation mapping observations of LILs. 

\end{enumerate}  

\begin{figure}[h!]
    \centering
    \includegraphics[width=0.75\textwidth]{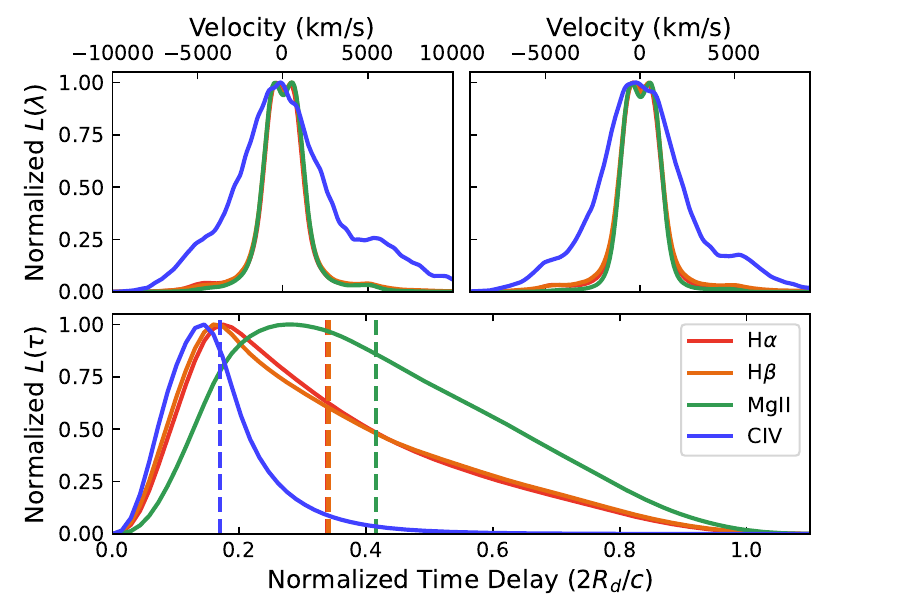}
    \caption{The mean line profiles (top-left), RMS line profiles (top-right), and 1DRFs (bottom) for a 2-zoned BLR model comprising a rotating disk with a biconical wind.  The dashed lines indicated the RWD for each line's response.  The velocity fields for the disk and bicone zones are Keplerian with turbulence and rotating outflows with turbulence, respectively.  The model parameters are listed in Table~\ref{tab:2zone:para} and the measured model values are listed in Table~\ref{tab:2zone:meas}. All 4 BELs are emitted in both zones. %The FWHM of the LILs is just about 2,000\,km\,s$^{-1}$. The FWHM of CIV is about 4,700\,km\,s$^{-1}$, whether the half-max is taken at 0.5 or 0.45.  The blue peak of CIV is at $-1,000$\,km\,s$^{-1}$.
    }
    \label{fig:2zone}
\end{figure}

In general, it is clear that the BLR response models cannot reproduce the observed differences mentioned above between HILs and LILs using only a single zone with a single-component velocity field.  To reconcile the observed differences, many authors have proposed two component BLR models including a rotating disk with a biconical outflow (i.e., a wind; e.g., \citet{Li2018Supermassive142,Yong2017TheModel,Elvis2000AQuasars}).  An example \textsc{Belmac} model of such a two-zone BLR is shown in Figure~\ref{fig:2zone}, with the parameters listed in Table~\ref{tab:2zone:para}. The disk zone has a Keplerian velocity field, clouds are concentrated towards the center, and have gas densities $(1-4.6)\times10^{11}$\,cm$^{-3}$.  The bicone zone is a rotating outflow, clouds are concentrated towards the center, and have a constant gas density, $n=10^{10}$\,cm$^{-3}$.  The velocity field of each zone also includes a turbulence component with FWHM $\approx300$\,km\,s$^{-1}$. The diagnostic parameters for this 2-zone model, including the FWHMs of the mean line profiles and the RWDs are listed in Table~\ref{tab:2zone:meas}.  The line ratios and equivalent widths are comparable to values derived from composite spectra from the SDSS \citep{VandenBerk2001CompositeSurvey} and the Large Bright Quasar Survey \citep{Francis1991ASpectrum}.

% LILs tend to be symmetrical and vary in the wings on short time-scales compared to the core:
The mean line profiles of the LILs have symmetrical shapes, with a double-peaked core, characteristic of a rotating disk. Their RMS profiles are also symmetrical, though MgII has slightly narrower wings than H$\alpha$ and H$\beta$.  The 1DRFs of the Balmer lines peak at $\tau\approx 0.2$, followed by a steady decline to zero. The 1DRF of MgII exhibits a much slower rise to a peak at $\tau\approx 0.3$ and a much more gradual decay.         
%Table~\ref{tab:2zone:meas} lists the mean line profile and 1DRF parameters for the 2-zone model, as well as the virial SMBH mass determined from the \textcolor{teal}{RWD and FWHM}.

% HILs tend to be broader, blue-asymmetric, and blueshifted with respect to the LILs:
The mean CIV profile is blue-asymmetric and broad, with a FWHM $\approx5,000$\,km\,s$^{-1}$, twice as large as the FWHM of the LILs ($\approx2,500$\,km\,s$^{-1}$). Determining if CIV is blueshfted with respect to the LILs (centered on $v^{||}=0$\,km\,s$^{-1}$) depends on how the ``center" of the profile is defined.  If the velocity shift is measured as the shift of the strongest peak relative to $v^{||}=0\,$km\,s$^{-1}$, CIV is blueshifted by about $-600$\,km\,s$^{-1}$.  However, within the uncertainty, there is no shift if we use the median velocity of the profile.  The 1DRF of CIV peaks at $\approx0.15$, then decreases more rapidly than for the LILs.  The CIV model has the shortest RWD of 0.17, about 60\% shorter than the RWD for MgII.  

\begin{table}[h!]
\centering
\caption{2-zone BLR model parameters for a rotating disk with a biconical wind.  The whole BLR has an integrated $C_f=0.30$, $M_\bullet=10^8\,M_\odot$, and $i=45^\circ$.} 
\label{tab:2zone:para}
    \begin{tabular}{|c||cc|}
    \hline
    \multirow{2}{*}{Parameter} & \multicolumn{2}{c|}{Zone Value} \\ \cline{2-3} 
     & \multicolumn{1}{c|}{Disk} & Bicone \\ \hline\hline
    Velocity Field\footnote{K = Keplerian, Ro = radial outflow, T = turbulence}& \multicolumn{1}{c|}{K+T} & K+Ro+T \\
    $v(R_{in})/v_{esc}$ & \multicolumn{1}{c|}{$0.5$} & $1$ \\ 
    Y & \multicolumn{1}{c|}{30} & 30 \\
    $\sigma$ & \multicolumn{1}{c|}{$25^\circ$} & $20^\circ$ \\
    $\mathcal{O}$ & \multicolumn{1}{c|}{$77.5^\circ$} & $50^\circ$ \\
    $\log\,(\frac{n(R_d)}{\mathrm{cm}^{-3}})$ & \multicolumn{1}{c|}{11.0} & 10.6 \\
    $C_{f,zone}$ & \multicolumn{1}{c|}{0.20} & 0.10 \\
    $p$ & \multicolumn{1}{c|}{0.5} & 0 \\
    $s$ & \multicolumn{1}{c|}{$-0.6$} & $0$ \\ \hline
    \end{tabular}
\end{table}

The asymmetric CIV profile shape and symmetric shapes of the LILs can be understood as follows:  1) Line opacity has a strong impact on profile asymmetries that can counter the asymmetry caused by the HICM. Therefore, optically thick lines (such as LILs) tend toward symmetrical profile shapes.  The column densities of clouds in the disk and bicone zones are $\sim10^{24}$\,cm$^{-2}$ and $\sim10^{23}$\,cm$^{-2}$, respectively, large enough to create optically thick clouds.  %The asymmetry caused by line opacity is balanced by the asymmetry caused by the inter-cloud medium, thus the profile shapes of the optically thick lines are more symmetric.  
CIV, an optically thin line, is less affected by ACE, therefore, its shape retains the blue-asymmetry caused by the HICM.  2) Most of the CIV flux is emitted at small radii in the biconical wind.  The column densities in the rotating disk zone are greater than the biconical wind %due to its overall higher gas density and cloud sizes.  
but, the ionization parameter in the disk, $<U>\approx0.001$, is lower than the bicone, $<U>\approx0.2$.  Therefore, CIV emission from the rotating disk contributes very little to the total CIV line emission.  %In the biconical wind zone, the ionization parameter is also constant with distance, but higher than the disk zone, $\log\,<U>\approx-2$.  Additionally, and unlike the disk-zone, the clouds are concentrated towards the center.  As a result, the emission from the inner region of the biconical zone is dominated by CIV. 

\begin{table}
\centering
\caption{Measured BEL properties for the 2-zoned BLR model.} 
\label{tab:2zone:meas}
    \begin{tabular}{|c||c|c|c|c|c|c|}
    \hline
    BEL & $L_\lambda/L_{H\beta}$ & EW$_\lambda$\footnote{The continuum luminosity is calculated from the \citet{Jin2012AProperties} composite SED used for the models.} & 50 Percentile & Peak(s) & FWHM & RWD \\ \hline\hline
    H$\alpha$ & $3.67$ & $98\AA$ & $42\pm100$\,km\,s$^{-1}$ & $\pm450\pm50\,$km\,s$^{-1}$ & $2,500\pm100\,$km\,s$^{-1}$ & $0.340$ \\
    H$\beta$ & 1.00 & $21\AA$ & $57\pm100$\,km\,s$^{-1}$ & $\pm450\pm50\,$km\,s$^{-1}$ & $2,500\pm100\,$km\,s$^{-1}$ & $0.345$ \\
    MgII & $1.94$ & $27\AA$ & $70\pm100$\,km\,s$^{-1}$ & $\pm450\pm50\,$km\,s$^{-1}$ & $2,500\pm100\,$km\,s$^{-1}$ & $0.416$ \\
    CIV & $2.61$ & $37\AA$ & $125\pm100$\,km\,s$^{-1}$ & $-600\pm50\,$km\,s$^{-1}$ & $5,000\pm100\,$km\,s$^{-1}$ & $0.172$ \\ 
    \hline
    \end{tabular}
\end{table} 

\subsection{Radial Flows and BEL Asymmetry}\label{sec:dis:asym}

In Section~\ref{sec:res:RDisk} we presented models for radial in- or outflow velocity fields, which display various types of asymmetry in the core and wings of their mean line profiles.  The types and degrees of asymmetry are consequences of the radial velocity field, line radiative transfer within clouds, and opacity due to the HICM. 

Blue-asymmetric and/or blueshifted profiles are commonly assumed to indicate the presence of an outflow. Line-emitting clouds flowing toward the observer produce blueshifted line emission. However, as explained in Section~\ref{sec:res:RDisk}, at the high gas densities and column densities thought to be characteristic of BLR clouds, internal line opacities are generally significant and therefore, clouds emit anisotropically in most lines, even, to some extent, in HILs like CIV. This effect (ACE) tends to suppress blueshifted emission from approaching clouds in the nearside of the BLR.  On the other hand, a distributed source of opacity, such as electron scattering in a HICM, will tend to suppress redshifted emission produced by receding clouds on the far side of the BLR. Without distributed, inter-cloud opacity, ACE actually results in red-asymmetric profiles for radial outflow velocity fields, and vice-versa for radial inflows.

\begin{figure}
    \centering
    \includegraphics[width=0.9\textwidth]{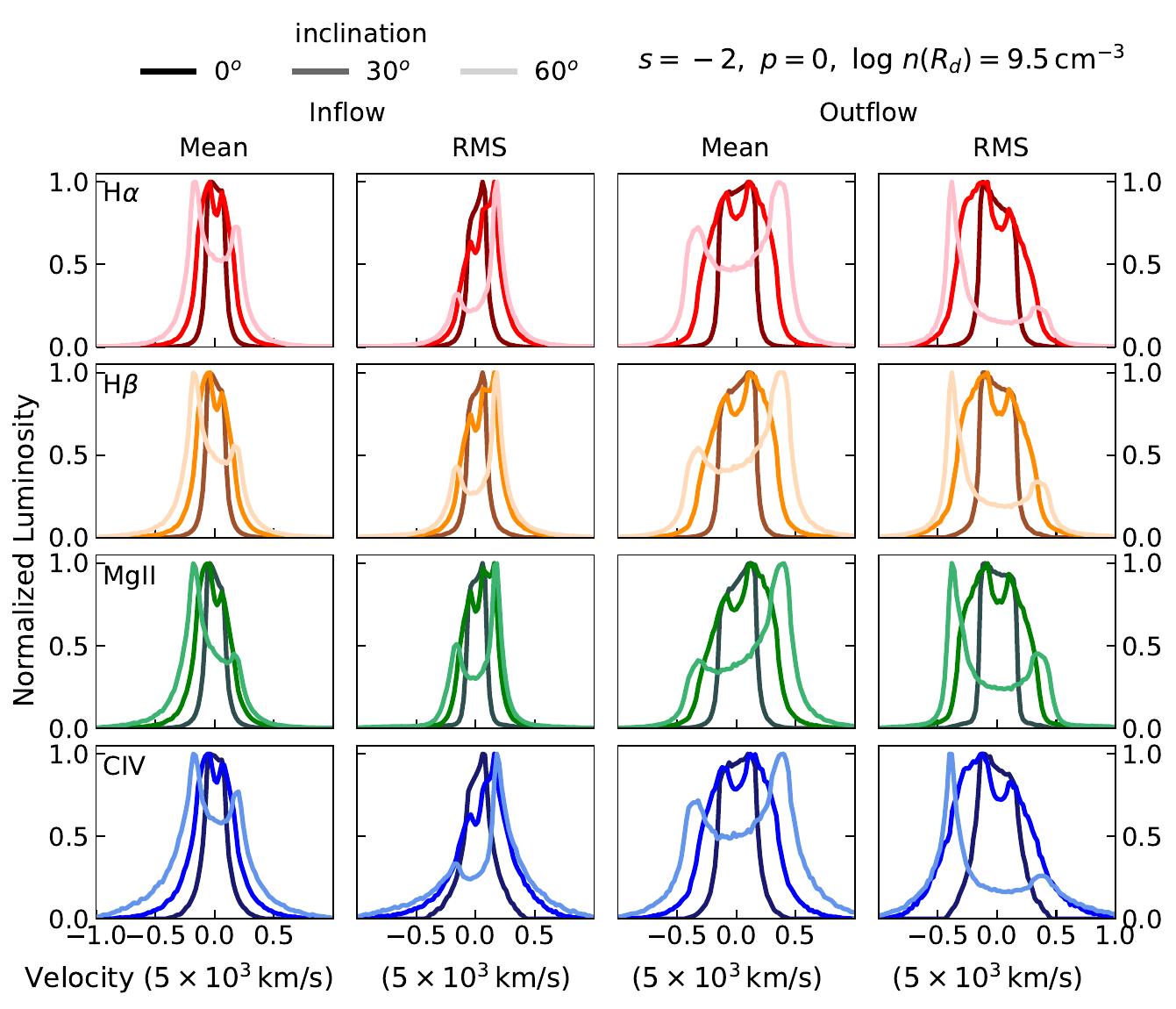}
    \caption{The same as Figure~\ref{fig:RDisk:s-2p0}, but for models without the hot, inter-cloud medium.  The asymmetry of the mean line profiles is the opposite of those in Figure~\ref{fig:RDisk:s-2p0}.}
    \label{fig:RDisk:s-2p0:noHICM}
\end{figure}

We demonstrate this in Figure~\ref{fig:RDisk:s-2p0:noHICM}, which shows the mean and RMS profiles of the same radial inflow and outflow models presented in Figure~\ref{fig:RDisk:s-2p0}, but without a HICM. The mean line profiles have the opposite asymmetries compared to those in Figure~\ref{fig:RDisk:s-2p0}. The clouds emit preferentially from their illuminated faces (the ACE effect) and therefore the inflow models produce blue-asymmetric line profiles whereas the outflow models produce red-asymmetric profiles. In contrast, the RMS profiles differ only slightly and maintain the same asymmetries as in Figure~\ref{fig:RDisk:s-2p0}. As the clouds in the nearside of the BLR respond with shorter delays than those in the far-side, the RMS profiles always have blue asymmetries for outflows and red asymmetries for inflows.

However, even if the HICM is sufficiently dense to cause blue-asymmetric profile cores (resulting in blueshifted peaks) when the velocity field is a radial outflow, our models suggest that the wings don't necessarily have the same type of asymmetry, depending on the radial velocity gradient. For example, in the case of a decelerating outflow, a relatively optically thin line such as CIV could have a profile with a blueshifted peak but an extended {\em red} wing (see Figure~\ref{fig:RDisk:asy}).  

A blue asymmetry in a given BEL could therefore be produced by either a radial outflow or inflow, depending on the optical depths of respectively, individual clouds (for that line), and the HICM. A radial inflow would produce the opposite asymmetries. However, this ambiguity does not exist for the RMS profiles. Indeed, comparing the RMS profiles with the core and wing asymmetries of the mean line profiles for different BELs could in principle provide diagnostics of the velocity field's flow direction, radial gradient, and the dominant opacity source.     

The density of the HICM that we have adopted, $n_e(r)=10^6\,$cm$^{-3}$, is consistent with previous estimates \citep{Netzer2008IonizedNuclei,Goncalves2001EvolutionLines,Reynolds1995WarmNuclei} and is about the lowest density needed to produce blue-asymmetric H$\alpha$ profiles for the models presented in this paper.  A higher electron density would increase the degree of asymmetry but would also make the BLR as a whole optically thick, which is inconsistent with observations. 
% Although we do not explore the effect of varying the inter-cloud medium density here, the total H$\alpha$ flux would be extinguished by 2 orders of magnitude if $n_e(r)=10^7\,$cm$^{-3}$.  The radial distribution of the inter-cloud medium will also affect the line profile asymmetry, that is, it will influence the comparative asymmetry between the wings and core of the profiles.  In the current version of \textsc{Belmac} as presented here, the density distribution is assumed to be uniform.  However, it is reasonable to assume that the inter-cloud medium density decreases with radius since, if mass is conserved, $n_e(r)\propto r^{-2}$.  If the inter-cloud medium density has a strong gradient with radius, emission from clouds near the center of the BLR could be suppressed more than the emission from the far-side clouds with longer LOS path lengths.  In this scenario, the flux ratio between the blue wing, red wing, and core of a profile would change with the radial velocity field in a way different from the picture described in Section~\ref{sec:res:RDisk} and Figure~\ref{fig:RDisk:asy}. 
% Alternative text
For example, the total H$\alpha$ flux would be extinguished by 2 orders of magnitude if $n_e(r)=10^7\,$cm$^{-3}$. In the \textsc{Belmac} models presented here, the density distribution of the HICM is assumed to be uniform. However, it is reasonable to assume that the density decreases with radius since, if the mass is conserved, $n_e(r)\propto r^{-2}$. The emission from clouds near the center of the BLR could then be suppressed more than the emission from the far-side clouds with longer LOS path lengths. This would result in wing and core asymmetries different from those described in Section~\ref{sec:res:RDisk} and illustrated in Figure~\ref{fig:RDisk:asy}.
We defer an exploration of the effects of varying $n_e(r)$ to future work.   

Measurements of both the velocity shift and asymmetry of a line profile will depend on defining the center of the line. 
For the models, we can simply define the center of the mean line profiles as $v^{||}=0\,$km\,s$^{-1}$, the zero of the LOS velocity. However, this is less clear in observed spectra.
%, especially in the absence of strong narrow lines to accurately establish the systemic velocity (as is usually the case for BELs in the rest-frame UV). 
The ``center'' of a BEL can be defined using the line's rest frame wavelength (assuming an accurate redshift is available), or the median wavelength of the line profile, or the line peak. As shown in Section~\ref{sec:dis:bels}, the line's peak or centroid does not necessarily coincide with the true zero of $v^{||}$.  For example, due to asymmetries in the profile core, the strongest peaks of the mean line profiles shown in Figure~\ref{fig:RDisk:s-2p0} are shifted relative to $v^{||}=0\,$km\,s$^{-1}$. The peaks of the  optically thin/moderately thick BELs (i.e., CIV and the Balmer lines) are blueshifted by $v^{||}\approx400$ to $\approx2,000\,$km\,s$^{-1}$ (with increasing inclination) for the outflow models and redshifted from $v^{||}\approx300$ to $\approx1,000\,$km\,s$^{-1}$ for the inflow models. 
%The clouds in the radial outflow models are launched from $R_{in}$ at a greater velocity than the inflow models, which causes the larger velocity shift of the profile peaks.  
The opposite behavior occurs for the optically thick MgII line, for which the profile peaks are blueshifted from $v\approx100$ to $\approx1,000\,$km\,s$^{-1}$ for the inflow models and redshifted from $v\approx500$ to $\approx2,000\,$km\,s$^{-1}$ for the outflow models (Figure~\ref{fig:RDisk:s-2p0}). 

Therefore, the choice of the line ``center'' can influence the perception and interpretation of the line asymmetries. For example, defining a blue-shifted peak as the line center would create the impression of an overall red asymmetry.

\subsection{Implications for Determining SMBH Masses} \label{sec:dis:mass}

\begin{figure}[b]
    \centering
    \includegraphics[width=0.9\textwidth]{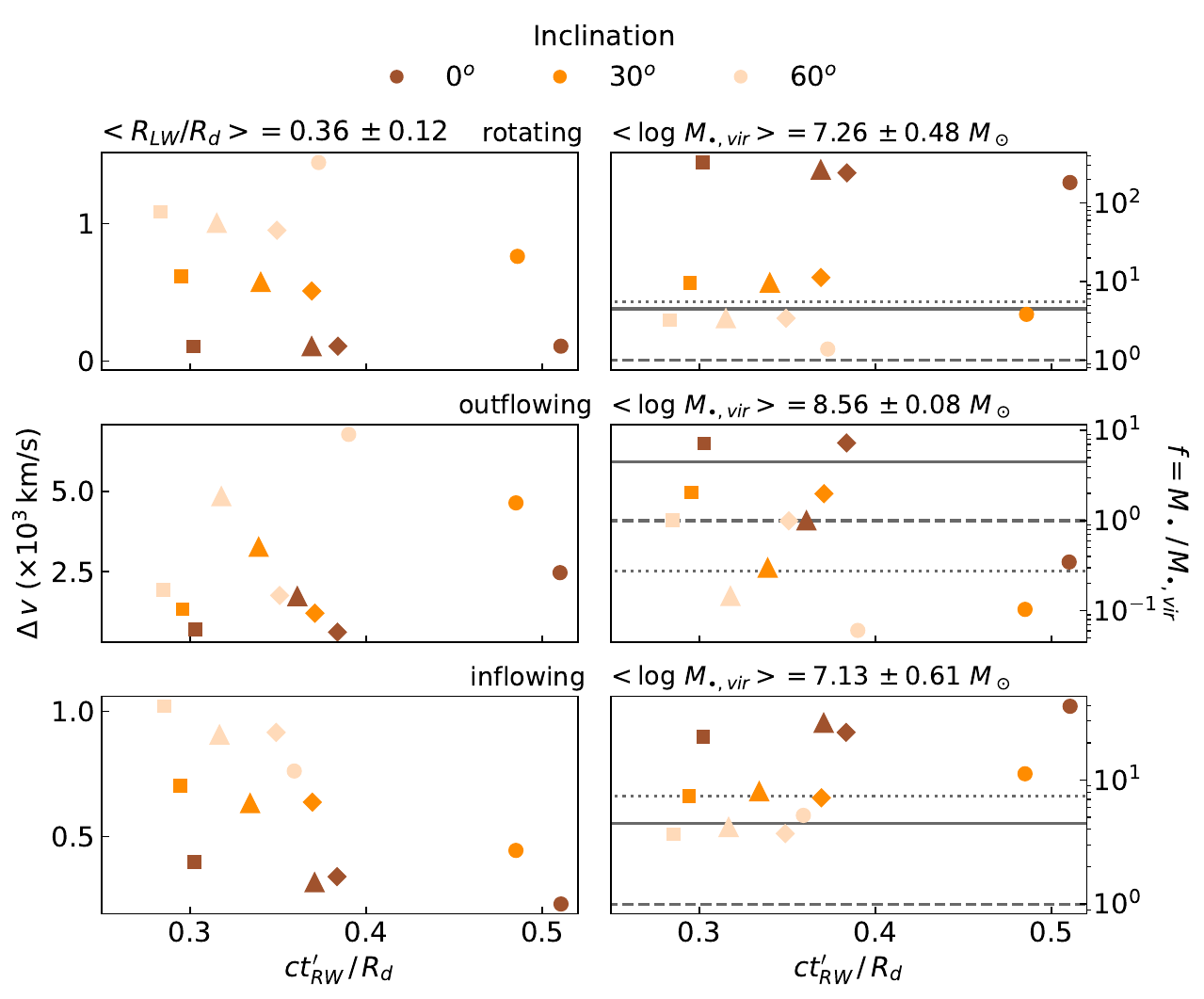}
    \caption{The BLR velocity dispersion ($\Delta v$; left), and virial factor ($f=M_\bullet/M_{\bullet,vir}$; right) plotted against the  BLR size inferred from the time delay (2\,RWD $=ct^\prime_{RW}/R_d=R_{LW}/R_d$) for the disk models of Sections~\ref{sec:res:KDisk} and~\ref{sec:res:RDisk}. The  H$\beta$ line was used to determine the values of $\Delta v$ and RWD. From top to bottom rows, the BLR models are rotating disks, radial outflows in a disk, and radial inflows in a disk.  The different shades represent different disk inclinations, as in Figure~\ref{fig:RDisk:s-2p0:noHICM}.  The symbols represent different $s$ and $p$ parameter combinations. Various $(s,\,p)$ combinations are represented by circles $(0,\,0)$, squares $(-2,\,0)$, triangles  $(0,\,2)$, and diamonds $(-2,\,2)$.  All the models have a true SMBH mass of $M_\bullet=10^8\,$M$_\sun$. The dotted line indicates the average value of $f$ for the models, the dashed gray line is $f=1$, and the solid gray line marks the commonly used empirical value, $f=4.5$.  The average virial SMBH mass and its RMS for each set of velocity field models is reported above the corresponding sub-plot.  The average luminosity-weighted radius, including all $Y_{BLR}=20$ models for H$\beta$, is $R_{LW}=0.36\pm0.12\,R_d$. 
    }
    \label{fig:mass}
\end{figure}

 As noted in Section~\ref{sec:intro}, SMBH masses in AGN are typically estimated using the virial theorem. The FWHM of a particular BEL is used as a proxy for the BLR velocity dispersion, $\Delta v$,  and the radius of the BLR, $R_{BLR}$, is obtained either directly from the reverberation lag or from the radius-luminosity relation. In addition, the virial factor, $f$, encapsulates the unknown specifics of the geometry, gas distribution, and velocity field. %Clearly, the value of $f$ will be sensitive to the properties of the BLR. 
 To illustrate the sensitivities of the virial factor to the properties of the BLR, we have calculated the virial product, $M_{\bullet,vir}$, in Equation~\ref{eq:mass} using RWD and FWHM values derived from the response models for a disk-like BLR with rotational, radial outflow, and radial inflow velocity fields (Sections~\ref{sec:res:KDisk} and~\ref{sec:res:RDisk}). We used the results for H$\beta$ as it is the BEL most commonly used in reverberation mapping. The velocity dispersion was obtained from the FWHM of the mean line profile ($\Delta v=\,$FWHM/2.355) and $R_{BLR}$ was calculated from the time delay corresponding to the RWD, $R_{BLR} = ct^\prime_{RW}$ (see Section~\ref{sec:results}). The corresponding value of $f$ was then evaluated as $f= M_\bullet/M_{\bullet,vir}$ for the actual SMBH mass used in the models, $M_\bullet=10^8$\,M$_\odot$. 
 
 In Figure~\ref{fig:mass}, $\Delta v$ (left column) and $f$ (right column) are shown as functions of $t^\prime_{RW}$ in units of BLR light crossing-time, for each of the 3 velocity fields.  In the panels of the right column, the solid lines mark the empirically determined value $f=4.5$.  The dashed and dotted lines indicate $f=1$ and the average value of $f$ obtained from the models, respectively. As both the cloud density distribution and the gas density distribution affect the RWD and hence the derived value of $R_{BLR}$ (see Section~\ref{sec:res:rwds}), the various combinations of $(s,~p)$ are indicated by different symbols. The time delay increases with the $(s,~p)$ combinations in the order $(-2,~0)$, $(0,~2)$, $(-2,~2)$ and $(0,~0)$. 

For all the velocity fields, $\Delta v$ increases as the inclination increases from $i=0^\circ$ to $60^\circ$. This causes correspondingly large variations in $f$ since, for a thin disk, the range in LOS velocity scales as $\sin i$ and therefore $M_{\bullet,vir}$ as $\sin^2 i$. The profile shapes and hence $\Delta v$ are also affected by the $(s,~p)$ combinations, but the dependence is relatively weak compared to the inclination effect. The virial factor also depends on the $(s,~p)$ combinations through $R_{BLR}$, but the effect is again relatively small compared to that of inclination.  The models that exhibit negative responses $(s=0,~p=0)$ have larger RWDs than their counterparts with the same inclination, which for the rotational and outflow velocity fields contributes to their higher values of $M_{\bullet,vir}$. In the case of the inflow models, however, the larger RWD is outweighed by the smaller values of $\Delta v$.  For the rotating and inflowing disks $f > 1$ in all cases, which implies that $M_{\bullet,vir}$ underestimates $M_{\bullet}$ for these velocity fields.  The $i=60^\circ$ models tend to fall close to the empirical value, with $f\approx 3-5$. For most of the $i=30^\circ$ models $f\approx 7-10$, while for the $i=0^\circ$ models, which have the smallest values of $\Delta v$, $f\approx 30 - 300$\footnote{We note that $i=0^\circ$ is a limiting case, which for these single zone disk models, does not produce realistic BEL profile widths (which are $\lesssim$ a few hundred\,km\,s$^{-1}$).}  

For the radial velocity fields, the variation in $\Delta v$ with $i$ is strongly dependent on the gas density distribution.  When $s=-2$, the clouds initially decelerate moving outward from the inner radius ($dv(r)/dr < 0$), whereas for $s = 0$ they accelerate (see \citetalias{Rosborough2024ModelingNuclei}, Figure 5). As a result, for the same values of $i$ and $p$, the line widths can differ by a factor of $\gtrsim 2$ between models with $s=0$ and $s=-2$. Consequently, $M_{\bullet,vir}$ is larger for $s=0$ models, resulting in $f < 1$ for all inclinations (in fact $f< 0.1$ for the $i=60^\circ$, $(0,~0)$ case). In contrast, for the $s=-2$ models, $f > 1$ for $i\leq 60^\circ$.  Although $v(r)$ also depends on the cloud properties for radial inflows, the initial infall speed is $v((R_d) = v_{Kep}(R_d)$, whereas the outflowing wind models have an initial launch velocity equal to the escape velocity, $v(R_{in}) = v_{esc}(R_{in}$) (see Section~\ref{sec:meth:new}). With these initial conditions, the clouds accelerate inward in all of the inflow models.  Overall, $f$ ranges from $\approx 5\times 10^{-2}$ to $\approx 7$ for the radial outflow models.  For $s=0$, in particular, $M_{\bullet,vir}$ overestimates $M_{\bullet}$ for all inclinations. In general, these models show that if the velocity field is dominated by an escaping outflow, such as in super-Eddington AGN, the virial theorem may significantly overestimate the SMBH mass.

\subsection{The Response Weighted Delay and the Radius -- Luminosity Relationship}\label{sec:dis:RL}

The average luminosity-weighted radius of the models used in Figure~\ref{fig:mass} is $R_{LW}=0.36\pm0.12\,R_d$ or about 173 light-days (ld), where most of the spread is due to the models with negative responses. For comparison, this is about $4\times$ greater than the value of $R_{BLR}$ implied by the $R-L$ relation defined by Equation 2 in \citet{Bentz2013THENUCLEI}, assuming a bolometric correction factor of 9 \citep[e.g.,][]{Netzer2019BolometricNuclei}. It seems likely that this discrepancy arises at least partly because the value of $R_d$ we have adopted overestimates the inner radius of the torus, and hence the outer radius of the BLR. It is well known that radii inferred from K-band reverberation mapping, which should sample the hottest dust in the innermost regions of the torus, are smaller by factors of 2--3 than implied by Equation~\ref{eq:Rd}, which is based on the standard interstellar medium (ISM) dust grain composition \citep[e.g.,][]{Kishimoto2007TheReverberation}. Torus reverberation modeling using \textsc{Belmac}'s predecessor code, \textsc{Tormac}, has been presented by \citet{Almeyda2020ModelingFunctions}. They similarly found that for the standard ISM dust composition, the model $R_{LW}$ is larger by a factor of 4 than the radii derived from the K-band $R-L$ relation. However, they also found that the discrepancy is reduced by a factor $\sim 2$ for models in which the torus is anisotropically illuminated by the accretion disk continuum. An alternative, or additional, explanation is that the K-band dust emission is mainly produced by large graphite grains, which have a higher sublimation temperature and therefore can survive inside the dust sublimation radius defined by Equation~\ref{eq:Rd} \citep{Goad2012TheTorus,Vazquez2015ConstrainingStudy,Honig2017DustyModels,Almeyda2020ModelingFunctions,Mor2012HotNuclei,Mor2009DUSTYDUST}. Although we have not explored anisotropic illumination in the \textsc{Belmac} models presented here, it would likely have a similar effect for a fixed value of $Y_{BLR}$. It may also be the case that the outer boundary of the BLR is better approximated by adopting a higher value of $T_{sub}$ in Equation~\ref{eq:Rd}, that is more representative of large graphite grains.  In reality, however, rather than a well-defined boundary between the BLR and the torus, there is likely to be a transition region where large graphite grains survive in the interiors of the photoionized BLR clouds \citep[e.g.,][]{Baskin2018DustNuclei}.  This region would produce both broad line and near infrared continuum emission, and will be incorporated in future versions of \textsc{Belmac}. %Another possible reason why $R_{BLR}$ inferred from the observed reverberation lags is smaller than $R_{LW}$ calculated from the \textsc{Belmac} models is the diffuse nebular continuum produced by the BLR. The nebular continuum emission (in the optical band, primarily the H and He recombination continua and the FeII ``pseudo continuum'') and BELs respond to ionizing UV variations on the same timescale.  Therefore, if the optical light curve, which is commonly used as a proxy for the driving (ionizing continuum) light curve, is significantly contaminated by the nebular continuum, the BLR cross-correlation delays will be biased towards shorter delays \citep{Korista2019QuantifyingDelays,Korista2001TheClouds}.  The BLR nebular continuum emission can be added to the photoionization model grids and its reverberation response can be modeled with \textsc{Belmac} in future work.  
Lastly, here we have used a square-wave pulse as the driving light curve in order to approximate the transfer function and assumed that the RWD is equivalent to the cross-correlation lag.  This is not necessarily the case, as the lag determined from cross-correlation analysis of real AGN light curves is affected by the cadence, observing epoch, and the overall duration of the observing campaign \citep{Perez1992Thesimulations, Goad2014InterpretingResponse}.  Therefore, realistic AGN light curves should be used for a more meaningful comparison between the model RWD values and the $R-L$ relation.  

In cases where the 1DRF exhibits at least a partly negative response, the RWD can overestimate the LWR by a factor 2 or more (LWR/RWD $\lesssim 0.5$, Figure~\ref{fig:Disk:rwd/lwr}).  In \citetalias{Rosborough2024ModelingNuclei} we reported a similar effect caused by the presence of matter-bounded clouds; as they are already fully ionized, the Hydrogen recombination line emission cannot increase in response to the continuum pulse, resulting in a longer RWD.  More generally for photoionized gas clouds, the line emission is strongly dependent on the ionization parameter.  Typically, the line flux increases with $U$ until it reaches a maximum, then decreases as $U$ increases further.  The value of $U$ at which the flux peaks differs between emission lines, however, it is approximately the value at which the Str\"omgren column density, $N_s\sim 10^{23}\,[\text{cm}^{-2}]\,U$ equals the cloud column density (e.g., in Figure~\ref{fig:FvU}, the line fluxes peak at $\log U \sim -1$ and $1$, respectively for models shown in the left and right panels). 
%This is analogous to the clouds becoming matter-bounded in the pure Hydrogen models investigated by \citetalias{Rosborough2024ModelingNuclei}. When negative responses occur, the effects are similar to those of matter-bounded clouds in the models of \citetalias{Rosborough2024ModelingNuclei}, and the RWD is skewed towards longer delays. 
When this value of $U$ is exceeded, a negative response occurs and the RWD is skewed toward longer delays. Thus, our RWD analysis implies that in situations where the line response is inversely correlated to the driving light curve, the measured time delay may overestimate the size of the BLR, leading in turn to an overestimate of the SMBH mass.  
Similar effects have previously been reported and extensively discussed in \citet{Goad1993ResponseNuclei, OBrien1994ResponseEmission}. %These works discussed the implications of negative responses on cross-correlation analysis \citep{Clavel1991StepsIUE,Peterson1991StepsWavelengths}.}% and note that this behavior was observed in NGC 5548 reverberation mapping data .}
In future work, we will conduct further analysis with a full continuum light curve rather than a single square-wave pulse to better understand the impact negative responses have on measured time delays and in turn, $M_{\bullet}$.

It is worth noting, however, that in the models considered in this paper, the gas density and hence also the ionization parameter and column density, are defined by simple power-laws and are therefore single-valued at a given $r$. Consequently, all clouds within a certain radius can become over-ionized and hence exhibit negative responses in various lines. This effect would be partly mitigated in configurations, such as in locally optimally emitting cloud model proposed by \citet{Baldwin1995LocallyLines} in which clouds at the same radius have a wide range of density and size. %

\section{Conclusion} \label{sec:concl}

In this paper, we introduced the new version of \textsc{Belmac}, first presented in \citetalias{Rosborough2024ModelingNuclei}, which includes photoionization model grids for multi-emission line reverberation mapping simulations.  We have presented a parameter exploration of the velocity-resolved reverberation response for H$\alpha$, H$\beta$, MgII, and CIV, to investigate differences in the responses of these lines and in particular, better understand observed differences between high and low ionization broad emission lines.  In the future, we will introduce an accompanying parameter estimation procedure for multi-BEL, reverberation modeling.  The main results of this work are summarized below.           

\begin{enumerate}    

    \item The observed differences between HILs and LILs cannot be reproduced with single-zone models. Instead, we find it necessary to introduce two separate regions with different velocity fields and geometries to explain the observed differences in response behavior and line profiles. We presented a 2-zone BLR model, similar to those previously discussed in several other works \citep[e.g.,][]{Gaskell1982AMotions,Proga2004DynamicsRadiation,Elvis2017QuasarWind,Yong2017TheModel}, consisting of a rotating disk with a biconical wind (Figure~\ref{fig:2zone}) as an example of a possible BLR configuration that can qualitatively explain the differences.
    
    \item Electron scattering in a hot, inter-cloud medium (HICM) attenuates the line emission of clouds furthest from the observer whereas anisotropic emission from individual clouds, due to the effects of line opacity, preferentially suppresses the observed emission from clouds located in the near side of the BLR. Together, these sources of opacity cause competing asymmetries that affect the line profile shapes. Although, in general, radial inflows and outflows tend to produce red-asymmetric and blue-asymmetric line profile shapes, respectively, interpretation is not straightforward.  An accelerating outflow (inflow) tends to have blue (red) -asymmetric wings and a decelerating outflow (inflow) tends to have red (blue) -asymmetric wings.  The degrees of asymmetry in both the core and the wings of the profile depend on the line opacity, the distribution of the clouds, and the velocity distribution.  It is worth noting that in the absence of inter-cloud opacity, such as scattering in the HICM, anisotropic cloud emission alone produces {\em red asymmetric} line profiles in outflow models (and vice versa for inflows). 
    %Radial inflows and outflows tend to produce red-asymmetric and blue-asymmetric line profile shapes, respectively.  An accelerating outflow (inflow) tends to have blue (red) -asymmetric wings and a decelerating outflow (inflow) tends to have red (blue) -asymmetric wings.  The degrees of asymmetry in both the core and the wings of the profile depend on the line opacity, the distribution of the clouds, and the velocity distribution.  It is worth noting that in the absence of inter-cloud opacity, such as scattering in the hot, inter-cloud medium, anisotropic cloud emission alone produces red asymmetric line profiles in outflow models (and vice versa for inflows). 

%    \item In certain circumstances the line emission responds inversely to the driving continuum light curve, decreasing as the continuum increases in strength. This occurs in models where the cloud column density is low and the ionization parameter is high in the inner region of the BLR. As a result, the BEL response is characterized by a minimum, rather than a maximum, at short time delays. 
    %, resulting in the BEL response function inversely responding at short time delays. 

 \item In general, the line fluxes emitted by individual clouds do not increase monotonically with the ionization parameter. Rather the line fluxes peak when the value of $U$ is such that the Str\"omgren column density is about the same as the cloud column density ($U\sim N_c/10^{23}\,\text{cm}^{-2}$). As found in previous studies \citep{Sparke1993DoesUp, Goad1993ResponseNuclei, OBrien1994ResponseEmission}, this can result in the BELs exhibiting a negative response to the driving continuum light curve. This occurs in models where the cloud column density is low and the ionization parameter is high in the inner region of the BLR. 
    
   % \item In general, when the line response is positively correlated with the driving light curve, the RWD for all BELs is equivalent to the LWR to within $20\%$, and within $10\%$ for H$\beta$. This suggests that the cross-correlation lag provides a reasonably accurate measure of $R_{LW}$ in this case. However, in situations where the line response is inversely correlated with the driving light curve over a significant portion of the BLR, the RWD is biased toward longer values, which can differ from the LWR by up to $\sim150\%$. Thus, our models imply that measured cross-correlation lags may overestimate $R_{LW}$ in such circumstances.

\item In general, for the single pulse driving light curve considered here, the RWD for all BELs is equivalent to the LWR to within $20\%$, and within $10\%$ for H$\beta$. This suggests that in the ideal case, the cross-correlation lag provides a reasonably accurate measure of $R_{LW}$. However, in cases where the line response is negative over a significant portion of the BLR, the RWD is biased toward longer values, which can differ from the LWR by up to $\sim150\%$.  Thus, our models imply that measured cross-correlation lags may overestimate $R_{LW}$ in such circumstances. It should also be noted that for observed AGN light curves, the cross-correlation lag will also be affected by such things as the sampling and time baseline covered by the light curves.

    \item The virial mass derived from the RWD and mean line profile FWHM of H$\beta$ can differ dramatically from the actual SMBH mass in the disk BLR models considered here. This implies a correspondingly wide variation in the virial factor. Not surprisingly, varying inclination has a drastic affect, since the velocity dispersion scales with $\sin{i}$ and hence $M_\bullet,\,vir$, with $\sin^2\,i$. However, the velocity field is also important. For the models with rotational and radial inflow velocity fields $M_\bullet,\,vir$ invariably underestimates $M_\bullet$, by factors $\sim 10-100$ at lower inclinations. However, for escaping outflows, $M_\bullet,\,vir$ may under or over estimate $M_\bullet$ by up to a factor of $\sim10$, depending on whether the outflow decelerates or accelerates to its terminal velocity.
    
    %Since the virial theorem assumes the velocity field of the BLR is dominated by Keplerian motion, the SMBH mass can be under and over estimated by up to a factor of $\sim10$ if the velocity field is primarily escaping outflows.  Additionally, the BLR orientation will alter the LOS velocity, and therefore $M_\bullet$, by $\sin^2\,i$.  If the velocity field is purely circular motion, for an inclination from $0\degr$ to $60\degr$, we found $M_\bullet$ was underestimated by factors of about 200 and 3, respectively.

    \item The average value of $R_{LW}$ for our models is about 170 ld, which is $4\times$ greater than expected from the BLR $R-L$ relation \citep{Bentz2013THENUCLEI}.  This difference is possibly due to a smaller outer BLR radius than the presumed dust sublimation radius.  In addition, 
    %in this paper we used a narrow, square-wave pulse as the driving light curve to generate response functions.
    %which are numerical approximations of the transfer functions.  
    it is possible that the RWD calculated from the response functions, which represent an idealized situation, does not accurately represent the cross-correlation lags obtained from observed line and continuum light curves.  A more thorough investigation of the the \textsc{Belmac} models in the context of the $R-L$ relation should be conducted with full, realistic AGN light curves.                  
  
\end{enumerate}

\begin{acknowledgments}
This paper is based upon work supported by the National Science Foundation under AARG/Grant No. 2009508. 

We thank the anonymous referee for constructive comments that have helped to improve the manuscript.
\end{acknowledgments}


\vspace{5mm}
%\facilities{HST(STIS), Swift(XRT and UVOT), AAVSO, CTIO:1.3m}


\software{  
          \textsc{Cloudy v23} by \cite{Gunasekera2023TheCloudy}, 
          \textsc{python v3.11}
          }

\bibliography{references}{}
\bibliographystyle{aasjournal}



\end{document}